\documentclass[prd,superscriptaddress,twocolumn]{revtex4}

\usepackage{mathrsfs,amsmath}
\usepackage{eurosym}
\usepackage{amssymb}
\usepackage{bbold}

\usepackage{latexsym,epsfig,epstopdf}
\usepackage{amssymb,amsmath,amsthm}

\usepackage{times}
\usepackage{color}

\usepackage{lmodern}
\usepackage[utf8]{inputenc}

\newcommand{\ket}[1]{\left | #1 \right\rangle}
\newcommand{\bra}[1]{\left \langle #1 \right |}

\newcommand{\Tr}{\mathrm{Tr}}
\newcommand{\braket}[2]{\left\langle #1|#2\right\rangle}
\newcommand{\proj}[1]{\ket{#1}\bra{#1}}

\usepackage{graphicx}

\usepackage{color}

\usepackage{braket}

\usepackage[mathscr]{euscript} 

\def\be{\begin{equation}}
\def\ee{\end{equation}}
\def\bea{\begin{eqnarray}}
\def\eea{\end{eqnarray}}

\begin{document}

\title{Generalised uncertainty relations from superpositions of geometries}

\author{Matthew J. Lake}
\email{matthew5@mail.sysu.edu.cn}
\email{mjlake@ntu.edu.sg}
\affiliation{School of Physics, Sun Yat-Sen University, Guangzhou 510275, China}
\affiliation{School of Physical and Mathematical Sciences, Nanyang Technological University, Singapore}

\author{Marek Miller}
\affiliation{School of Physical and Mathematical Sciences, Nanyang Technological University, Singapore}
\affiliation{Instytut Fizyki Teoretycznej, Uniwersytet Wroclawski, pl. M. Borna 9, 50-204 Wroclaw, Poland}

\author{Ray F. Ganardi}
\affiliation{School of Physical and Mathematical Sciences, Nanyang Technological University, Singapore}

\author{Zheng Liu}
\affiliation{School of Physical and Mathematical Sciences, Nanyang Technological University, Singapore}
\affiliation{Centre for Quantum Technologies, National University of Singapore, Singapore}

\author{Shi-Dong Liang}
\affiliation{School of Physics, Sun Yat-Sen University, Guangzhou 510275, China}

\author{Tomasz Paterek}
\affiliation{School of Physical and Mathematical Sciences, Nanyang Technological University, Singapore}
\affiliation{MajuLab, International Joint Research Unit UMI 3654, CNRS, Universit\'e C\^ote d'Azur, Sorbonne Universit\'e, National University of Singapore, Nanyang Technological University, Singapore}

\begin{abstract}

Phenomenological approaches to quantum gravity implement a minimum resolvable length-scale but do not link it to an underlying formalism describing geometric superpositions.
Here, we introduce an intuitive approach in which points in the classical spatial background are delocalised, or `smeared', giving rise to an entangled superposition of geometries. 
The model uses additional degrees of freedom to parameterise the superposed classical backgrounds. 
Our formalism contains both minimum length and minimum momentum resolutions and we naturally identify the former with the Planck length. 
In addition, we argue that the minimum momentum is determined by the de Sitter scale, and may be identified with the effects of dark energy in the form of a cosmological constant.  
Within the new formalism, we obtain both the generalised uncertainty principle (GUP) and extended uncertainty principle (EUP), which may be combined to give an uncertainty relation that is symmetric in position and momentum.
Crucially, our approach does not imply a significant modification of the position-momentum commutator, which remains proportional to the identity matrix.  
It therefore yields generalised uncertainty relations without violating the equivalence principle, in contradistinction to existing models based on nonlinear dispersion relations.
Implications for cosmology and the black hole uncertainty principle correspondence are briefly discussed, and prospects for future work on the smeared-space model are outlined.

{\textbf{Keywords}: generalised uncertainty principle, extended uncertainty principle, minimum length, minimum momentum, quantum gravity} 

\end{abstract}

\maketitle
\tableofcontents

\section{Introduction}

Heisenberg's uncertainty principle (HUP) forbids simultaneous knowledge of both position and momentum to arbitrary precision, 
\begin{eqnarray} \label{HUP-1}
\Delta x \, \Delta p \gtrsim \frac{\hbar}{2} \, .
\end{eqnarray}  
It can be introduced using the Heisenberg microscope thought experiment \cite{aHeisenberg:1927zz,Heisenberg:1930}, in which irremovable uncertainty is explained heuristically as the result of momentum transferred to a massive particle by a probing photon, or derived from the quantum formalism, as shown by the pioneering work of Robertson \cite{Robertson:1929zz} and Schr\"odinger \cite{Schrodinger1930,Schrodinger:1930ty}.
In the latter, it is seen to arise from the general inequality
\begin{eqnarray} \label{Robertson-Schrodinger-1}
\Delta_\psi O_1 \, \Delta_\psi O_2 \geq \frac{1}{2} |\braket{\psi | [\hat{O}_1,\hat{O}_2] | \psi}|  \, ,
\end{eqnarray}  
where the uncertainty of the observable $\hat{O}$ is defined as the standard deviation 
\begin{equation}
\Delta_\psi O := \sqrt{ \braket{\psi | \hat{O}^2 | \psi} - \braket{\psi | \hat{O} | \psi}^2 },
\end{equation}
together with the non-commutativity of position and momentum,
\begin{eqnarray} \label{[x,p]-1}
[\hat{x},\hat{p}] = i\hbar \ \hat{\openone} \, .
\end{eqnarray}
A more careful statement of the HUP, derived from the underlying formalism of quantum mechanics (QM), therefore reads
\begin{eqnarray} \label{HUP-2}
\Delta_\psi x \, \Delta_\psi p  \geq \frac{1}{2} |\braket{\psi | [\hat{x},\hat{p}] | \psi}| = \frac{\hbar}{2} \, ,
\end{eqnarray}
where $\Delta_\psi x$ and $\Delta_\psi p$ are well defined, unlike the heuristic uncertainties $\Delta x$ and $\Delta p$ used in Eq. (\ref{HUP-1}).
 
More recently, the Heisenberg microscope argument has been generalised to include the gravitational interaction between the particle and the photon and in this way motivate the generalised uncertainty principle (GUP),  
\begin{eqnarray} \label{GUP-1*}
\Delta x \gtrsim \frac{\hbar}{2\Delta p} + \alpha \frac{G}{c^3}\Delta p \, , 
\end{eqnarray} 
where $\alpha \sim \mathcal{O}(1)$ \cite{Adler:1999bu,Scardigli:1999jh}. 
Unlike the HUP, which treats position and momentum on an equal footing, the GUP implies a minimum position uncertainty, of the order of the Planck length, but no minimum momentum uncertainty.

Applying similar arguments in the presence of an asymptotic de Sitter space background, in which the minimum scalar curvature is of the order of the cosmological constant $\Lambda$, it has been argued that the momentum uncertainty is modified such that 
\begin{eqnarray} \label{EUP-1}
\Delta p \gtrsim \frac{\hbar}{2\Delta x} + \eta \hbar\Lambda \Delta x \, , 
\end{eqnarray} 
where $\eta \sim \mathcal{O}(1)$ \cite{Bolen:2004sq,Park:2007az,Bambi:2007ty}. 
This relation, known as the extended uncertainty principle (EUP) \cite{Park:2007az,Bambi:2007ty}, implies a minimum momentum uncertainty of the order of the de Sitter momentum $\sim \hbar\sqrt{\Lambda}$, but no minimum position uncertainty. 
Thus, taking the GUP (\ref{GUP-1*}) and EUP (\ref{EUP-1}) together reintroduces position-momentum symmetry in the gravitationally-modified uncertainty relations.

More precise formulations of the GUP and EUP, which are consistent with the Robertson-Schr{\" o}dinger relation (\ref{Robertson-Schrodinger-1}), may be obtained by modifying the canonical Heisenberg algebra (\ref{[x,p]-1}) such that 
\begin{eqnarray} \label{modified_commutator-1}
[\hat{x},\hat{p}] = i\hbar(\hat{\openone} + \tilde{\alpha} \hat{x}^2 + \tilde{\eta} \hat{p}^2) \, ,
\end{eqnarray} 
where $\tilde{\alpha}$ and $\tilde{\eta}$ are appropriate dimensionful constants \cite{Kempf:1996ss}. 
Equations (\ref{Robertson-Schrodinger-1}) and (\ref{modified_commutator-1}) imply an uncertainty relation, also called the extended generalised uncertainty principle (EGUP) \cite{Park:2007az,Bambi:2007ty}, that contains quadratic terms in both position and momentum, and which reduces to both the GUP and the EUP in appropriate limits. 
However, in canonical quantum mechanics, the momentum operator may be identified with the Galilean shift-isometry generator of flat Euclidean space, up to a factor of $\hbar$ \cite{Ish95}. 
Similarly, the canonical position operator may be identified with the shift-isometry generator in Euclidean momentum space. 
Hence, modifications of the form (\ref{modified_commutator-1}) imply (a) modification of the symmetry group that characterises the background geometry on which the wave function $\psi(x)$ is defined, (b) modification of the canonical de Broglie relation, $p = \hbar k$, or (c) both. 

The consistency of these results with heuristic arguments for the existence of minimum length- and momentum scales suggests that, whatever their origin, modified commutation relations of the form (\ref{modified_commutator-1}) correctly capture certain aspects of quantum gravity phenomenology. 
Modified commutators have also been motivated by arguments invoking string theory, black hole physics, space-time non-commutativity and deformed special relativity, among others \cite{Tawfik:2015rva,Tawfik:2014zca}. 
Nevertheless, it remains unclear in what way (if any) such commutators are related to superpositions of classical geometries.
We recall that such superpositions are required by any self-consistent theory of quantum gravity, in which the principles of quantum mechanics, including quantum superposition, and general relativity, including gravity as space-time curvature, both hold. For brief but pertinent discussions on the necessity of quantising the gravitational field,
see \cite{Hossenfelder:2012jw,Garay:1994en} and references therein. 
For counter-arguments \cite{Carlip:2008zf,KTM14} and additional recent work, see \cite{Mauro15,Tanjung17,Marletto:2017pjr,Marletto:2017kzi,Bose:2017nin,Christodoulou:2018cmk,Rovelli2,Belenchia:2018szb}.

Here, we present a formalism that gives rise to both the GUP and EUP and, hence, to a modified uncertainty principle that is symmetric in position and momentum, which also reduces to the EGUP in a suitable limit.
However, contrary to many other approaches, we do not begin by modifying the canonical commutation relation, but seek a mathematical structure that permits quantum superpositions of classical geometries.  
The description of such superpositions is necessary if quantum particles are to act as sources of the gravitational field.

To this end, we think of points in physical space as quantum mechanical objects that can be described by vectors in a Hilbert space. 
We argue that, in $d$ spatial dimensions, the simplest way of representing such superpositions is via a quantum state with an additional $d$ degrees of freedom.
In this way, superpositions of one-dimensional geometries may be depicted, heuristically, using a two-dimensional plane. 
Similarly, superpositions of $d$-dimensional geometries, for $d \geq 2$, may be represented in a $2d$-dimensional hyper-plane.
Applying the same principle to the momentum space representation of the quantum state implies a doubling of the number of dimensions vis-{\` a}-vis the classical phase space of the theory.

The formalism introduced in this way contains $d$ free parameters that quantify the smearing of a classical point in each dimension of physical (position) space. 
These also represent the minimum uncertainties of position measurements in each of the $d$ coordinate directions. 
We naturally identify the minimum uncertainties associated with dimensionful coordinates (i.e.,  spatial coordinates with dimensions of length) with the $D$-dimensional Planck length, where $D = d+1$ is the number of space-time dimensions. 
However, we restrict our attention to the non-relativisitc limit and do not attempt to `smear' space-time in the present work.

An additional $d$ parameters quantify the smearing of points in classical momentum space, and also represent the minimum uncertainties of momentum measurements in each spatial direction.
Assuming the $\Lambda \rm CDM$ concordance model of contemporary cosmology \cite{Ostriker:1995rn,Reiss1998,Perlmutter1999,Betoule:2014frx,PlanckCollaboration}, we argue that the smearing scale in three-dimensional momentum space should be set by the de Sitter momentum $\sim \hbar\sqrt{\Lambda}$. 
The model is then formally extended to an arbitrary number of dimensions by substituting the $D$-dimensional cosmological constant, $\Lambda_{D}$.

We note that, for $d \geq 4$, $d-3$ spatial dimensions must be compactified \cite{Font:2005td}, or highly warped \cite{Maartens:2010ar}, in order to give rise to the $(3+1)$-dimensional universe we observe. 
The minimum momentum-scale may then be related to the size or warping of the extra dimensions \cite{Dupays:2013nm,Cline:2000ky}. 
However, we neglect such subtleties in our present analysis. 
Hence, we develop the smeared-space formalism for an arbitrary number of spatial dimensions, as a purely formal result, but focus on the three-dimensional case when discussing applications and possible observational consequences.

It is appealing that the resulting formalism links the existence of minimum length- and momentum-scales with the superposition of classical geometries.
However, it should be mentioned that, though many approaches to quantum gravity indicate the existence of a minimum length-scale, this is not universally accepted as a feature that any candidate theory must possess \cite{Rovelli:2000aw}.
We conclude this introduction with a brief review of current approaches to quantum gravity.
These include string theory \cite{Kiritsis:2007zza}, loop quantum gravity (LQG) \cite{Ashtekar:2012np}, asymptotically safe gravity \cite{Niedermaier:2006ns}, Euclidean quantum gravity \cite{EuclideanQG}, causal set theory \cite{Dowker:aza}, causal dynamical triangulations \cite{Ambjorn:2004qm} and group field theory \cite{Freidel:2005qe}, among others \cite{Rovelli:2000aw}. 
In addition, studies of light-cone fluctuations, which may be interpreted as superpositions of space-time geometries if quantised, are similar in spirit to our present work \cite{Hossenfelder:2012qg,Ford:1994cr,Ford:1996qc,Ford:1997zb}.  

The paper is organised as follows.
We first outline, in Sec.~\ref{SEC_SIMPLEST}, why the simplest approach to the smearing of classical points does not provide a consistent theory.  
We then introduce the consistent formalism in Sec.~\ref{SEC_FORMALISM}.
In Sec.~\ref{smeared_states}, we show how to smear canonical quantum states, and how to generalise canonical quantum operators to act on the smeared-states of the new formalism. 
Sec.~\ref{smeared_ops} then presents an alternative picture, in which the effects of smearing are incorporated into the definitions of observables, which continue to act on canonical quantum states. 
We show that both formulations of the smeared-space theory yield the same physical predictions, so that the two approaches may be thought of as roughly analogous to the Heisenberg and Schr{\" o}dinger pictures in canonical QM. 
In Sec.~\ref{SEC_UR}, we derive the unified uncertainty relation and discuss the limits in which the GUP and EUP are obtained independently. 
We give a basic outline of smeared-space wave mechanics in Sec.~\ref{SEC_WM}, and conclude our treatment of the formalism with a description of multi-particle states in Sec.~\ref{MULTI}. 
In Sec.~\ref{SEC_APPLICATIONS}, we consider possible applications of the theory, focussing on the implications of the generalised uncertainty relations for cosmology and black hole physics. 
Possible modifications of the smeared-space model due to finite-horizon effects are also considered. 
We note that, compared to the precise mathematical statements that define the formalism presented in Sec.~\ref{SEC_FORMALISM}, the arguments presented in this section are, necessarily, more speculative in nature. 
A summary of our results is given in the Conclusions, Sec.~\ref{CONCLUSIONS}, and prospects for future work are outlined. 
Finally, we show that generalised uncertainty relations can also be derived within the canonical quantum formalism, using an effective model in which `position' and `momentum' observables are described by positive operator valued measures (POVMs) \cite{Chuang_Nielsen} with finite resolution.  
We stress, however, that in such an effective model there are no fundamental limits on the uncertainties of position or momentum measurements, in contradistinction to those derived in the smeared-space formalism. 
These results are presented in the Appendix.

\section{Failure of the simplest idea} \label{SEC_SIMPLEST}

Consider a $d$-dimensional Euclidean universe, described by the manifold $\mathbb{R}^{d}$ equipped with the standard Euclidean metric.  
In the canonical quantum formalism, a point $\vec{x} \in \mathbb{R}^d$ in the classical background geometry may be represented, heuristically, by a $d$-dimensional Dirac delta wave function $\delta^d(\vec{x}{\, '}-\vec{x})$ or a ket $\ket{\vec{x}}$ in the Hilbert space of the theory
\footnote{
Strictly, $\delta^d(\vec{x}{\, '}-\vec{x})$ is the wave function of a particle that is located at the point $\vec{x}$ with 100\% certainty. 
In canonical QM, $\vec{x}$ labels a fixed point in the classical background geometry on which $\psi(\vec{x})$ is defined. 
Thus, $\ket{\vec{x}}$ represents the quantum state of an ideally localised particle defined on a fixed backgound, not the quantum state of a spatial point per se.}.
A quantum model of a smeared-space background is obtained by replacing this point by a
coherent superposition of all points in $\mathbb{R}^d$. 
This is most naturally realised by the map 
\begin{equation} \label{map-1}
\ket{\vec{x}} \mapsto \ket{g_{\vec{x}}} := \int g(\vec{x}{\, '}-\vec{x}) \ket{\vec{x}{\, '}} {\rm d}^{d}\vec{x}{\, '} \, ,
\end{equation} 
where the square of the `smearing function' $g(\vec{x}{\, '}-\vec{x})$ may be thought of as a $d$-dimensional Gaussian, 
whose width in each coordinate direction $x^{i}$, denoted $\sigma_{g}^{i}$, is assumed to be a fundamental property of quantum mechanical space. 
Here, ${\rm d}^{d}\vec{x}$ is shorthand notation for the volume element of classical position space, which includes the Jacobian given by the square root of the determinant of the classical metric.
  
Note that in the limit $\sigma_g^i \rightarrow 0$ for all $i$  
\footnote{Note the factor of $\sqrt{2}$ in the corresponding limit for the probability amplitude: $\lim_{\sqrt{2}\sigma_g^{i} \rightarrow 0}g(\vec{x}{\, '}-\vec{x}) = \delta^d(\vec{x}{\, '}-\vec{x})$.},
\begin{equation} \label{unsmeared_limit}
\lim_{\sigma_g^{i} \rightarrow 0} |g(\vec{x}{\, '}-\vec{x})|^2 = \delta^d(\vec{x}{\, '}-\vec{x}) \, , 
\end{equation} 
we recover the canonical theory, as Eq. (\ref{map-1}) maps each ket $\ket{\vec{x}}$, corresponding to a unique point $\vec{x} \in \mathbb{R}^d$, to itself. 
In general, an arbitrary quantum state, 
\begin{equation} 
\ket{\psi} = \int \psi(\vec{x}) \ket{\vec{x}} {\rm d}^d\vec{x} \, , 
\end{equation} 
is mapped according to
\begin{equation} \label{post-map}
\ket{\psi} \mapsto \int (g*\psi)(\vec{x}{\, '})\ket{\vec{x}{\, '}} {\rm d}^d\vec{x}{\, '} \, , 
\end{equation} 
where the star denotes a convolution.

In order to generate valid probabilistic predictions, the state (\ref{post-map}) must be normalised, independently of the original state $\ket{\psi}$.  
It is straightforward to demonstrate that this is possible if and only if $g(\vec{x}{\, '}-\vec{x})$ is a Dirac delta function.  
In this case, however, physical space is not smeared and remains classical. 
We must therefore consider alternative models of quantum-mechanically smeared space. 
The main idea of this paper, which is presented in the next section, is to introduce additional degrees of freedom that parameterise quantum fluctuations from the Euclidean background geometry, where the latter corresponds to the most probable quantum state. 
Using these, we are able to overcome the limitations of the simplest idea, presented above, to obtain a fully normalisable theory in the presence of smearing.

\section{Formalism} \label{SEC_FORMALISM}

\subsection{The smeared-state picture} \label{smeared_states}

\subsubsection{Smeared states}

\begin{figure}[!t] 
	\centering
	\includegraphics[width=8.7cm]{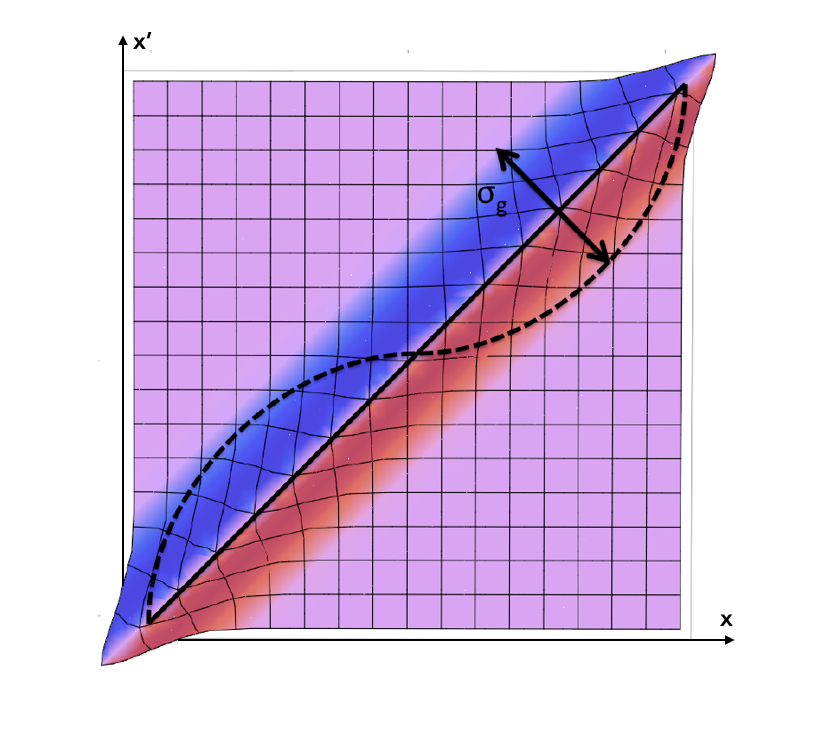}
	\caption{\label{FIG_PHASESPACE}
	Graphical representation of the plane associated with the smeared classical line. 
	The variable $x$, which labels spatial points in the one-dimensional Euclidean geometry, is plotted on the horizontal axis. 
	The variable $x'$, plotted on the vertical axis, parameterises the smearing of each point $x \in \mathbb{R}$. 
	Equivalently, it parametrises the non-local influence on $x$ from all points $x' \neq x$, as well as the influence of $x$ on itself when $x'=x$. 
	Thus, each point in the plane, $(x,x') \in \mathbb{R}^2$, is associated with a complex number $g(x'-x)$ that represents the amplitude for the transition $x \mapsto x'$. 
	In this example, the function $|g(x'-x)|^2$ is assumed to be a Gaussian centred at $x' = x$, with standard deviation $\sigma_g$, so that the straight black line represents the most probable one-dimensional universe. 
	This is isomorphic to the original classical geometry. 
	However, the curved dashed-black line represents a relatively probable geometry, in which spatial fluctuations remain within one standard deviation, $\sigma_g$, of the most probable (Euclidean) configuration. 
	We naturally identify this with the Planck scale, so that $\sigma_g \simeq l_{\rm Pl}$.}
\end{figure}

Let us again consider smearing a single point $\vec{x} \in \mathbb{R}^d$ with the smearing function $g(\vec{x}{\, '}-\vec{x})$.  
For fixed values of $\vec{x}$ and $\vec{x}{\, '}$, we interpret $g(\vec{x}{\, '}-\vec{x})$ as a quantum probability amplitude for the transition $\vec{x} \mapsto \vec{x}{\, '}$.  
Since, for each coordinate $x^i$, this involves a continuous parameter $x'^{i}$, the transitions are naturally represented within a $2d$-dimensional space, where each pair $(\vec{x},\vec{x}{\, '})$ is assigned the transition amplitude $g(\vec{x}{\, '}-\vec{x})$.
Fixed values of $\vec{x}$ correspond to parallel $d$-dimensional Euclidean hyper-planes and the $2d$-dimensional space emerges when we apply the smearing function to all classical points. 
This is illustrated, heuristically, for a toy one-dimensional universe, in Fig.~\ref{FIG_PHASESPACE}.

Let us repeat that the kets $\ket{\vec{x}}$ are the analogues of classical points $\vec{x}$ in the canonical quantum formalism.
Orthogonal directions in physical space are therefore represented by tensor products of the relevant Hilbert spaces. 
Hence, the Cartesian product between scalars corresponds to the tensor product between vectors, yielding the following correspondence between the classical and quantum phase spaces:
\begin{eqnarray} 
&&\vec{x} \leftrightarrow \ket{\vec{x}} \, , \ \ {\rm d}^d\vec{x} \leftrightarrow \ket{\vec{x}}{\rm d}^d\vec{x} \, , \ \ ( \ . \ , \ . \ ) \leftrightarrow . \otimes .
\end{eqnarray}

We therefore propose the following map as our model for the quantum smearing of a spatial point in the fixed-background theory:
\begin{equation}  \label{EQ_POINTMAP} 
\ket{\vec{x}} \mapsto \ket{\vec{x}} \otimes \ket{g_{\vec{x}}} \, ,
\end{equation} 
where $\ket{g_{\vec{x}}}$ is defined in Eq. (\ref{map-1}). 
The quantum state $\ket{g_{\vec{x}}}$ of the new (primed) degrees of freedom parametrises the spread of the original classical point $\vec{x}$.  
Equivalently, it parametrises the non-local influence, on $\vec{x}$, from all points $\vec{x}{\, '}$ in the classical background. 
In this way, we avoid the problem encountered in Sec.~\ref{SEC_SIMPLEST}, since an arbitrary state $\ket{\psi}$ in canonical quantum theory is now mapped according to: 
\begin{eqnarray} \label{EQ_PSIG}
\ket{\psi} \mapsto \ket{\Psi} &:=& \int\int g(\vec{x}{\, '}-\vec{x}) \psi(\vec{x}) \ket{\vec{x}} \otimes \ket{\vec{x}{\, '}} {\rm d}^d\vec{x} {\rm d}^d\vec{x}{\, '}
\nonumber\\
&:=& \int \psi(\vec{x}) \ket{\vec{x}} \otimes \ket{g_{\vec{x}}} {\rm d}^d\vec{x} \, .
\end{eqnarray}
Here, we introduce the shorthand ${\rm d}^d\vec{x} = \sqrt{\det g_{ij}(x)} {\rm d}^dx$, ${\rm d}^d\vec{x}{\, '} = \sqrt{\det g'_{ij}(x')} {\rm d}^dx'$, 
where $g_{ij}(x)$ and $g'_{ij}(x')$ denote the metrics on the subspaces defined by the conditions $\vec{x}{\, '} = {\rm const.}$ and $\vec{x} = {\rm const.}$, respectively. 

It is straightforward to show that $\ket{\Psi}$ is normalised for any normalised function $g(\vec{x}{\, '} - \vec{x})$.
In general, we denote by capital letters, e.g. $\ket{\Psi}$, the states and operators of the smeared-space model, 
and with lower case letters, e.g. $\ket{\psi}$, the states and operators of canonical QM. 
Physical predictions are assumed to be those of the smeared-space theory and the canonical QM of the original (unprimed) degrees of freedom is only a convenient tool in the smeared-space calculations.
We note that an arbitrary canonical state $\ket{\psi}$ is mapped to an entangled state $\ket{\Psi}$ in the tensor product Hilbert space.

In Fig.~\ref{FIG_PHASESPACE}, we illustrate the two-dimensional plane with which we visualise the smeared classical line.  
This represents a toy one-dimensional universe, which, though not physically realistic, helps us to visualise the smearing procedure (\ref{EQ_POINTMAP}). 
In the simplest scenario the square of the smearing function $|g(\vec{x}{\, '}-\vec{x})|^2$ is chosen to be a Gaussian, centred at $\vec{x}{\, '} = \vec{x}$, with standard deviation $\sigma_g^i$, $i \in \left\{1,2, \dots ,d \right\}$, in each of the $d$ spatial dimensions.
However, our proofs below hold for an arbitrary normalised smearing function, unless explicitly stated otherwise.
For smearing functions $g(\vec{x})$ with peak absolute values at $\vec{x}=0$, $\vec{x}{\, '}=\vec{x}$ remains the most probable value for each point, but deviations from the average within one standard deviation in each spatial direction are relatively likely.
In this way, the model contains superpositions of classical geometries, each of which is represented as a $d$-dimensional slice of the $2d$-dimensional space.

Thus, in our one-dimensional example, illustrated in Fig.~\ref{FIG_PHASESPACE}, the most probable geometry (solid line) is isomorphic to the original classical geometry and is simply the one-dimensional Euclidian universe.  
Parallel diagonal lines also represent Euclidean geometries, corresponding to situations in which each point in the classical line undergoes a transition $x \mapsto x' = x + a$, where $a$ is a constant.
Any other possible geometry is represented by a curve $x'(x)$ within the two-dimensional plane, e.g., the dashed curve also illustrated in Fig.~\ref{FIG_PHASESPACE}.  

In the general formalism, the induced metric on an arbitrary $d$-dimensional sub-manifold, defined by the vector function $\vec{x}{\, '}(\vec{x})$, may be obtained by performing the push-forward \cite{push-forward} from the metric on the $(\vec{x},\vec{x}{\, '})$-plane. 
Since each point $(\vec{x},\vec{x}{\, '})$ is associated with a quantum probability amplitude, this, in principle, allows us to calculate the amplitude associated with an arbitrary fluctuation away from the classical background geometry. 
However, a detailed investigation of this possibility lies outside the scope of the present paper.
The possible form of the $2d$-dimensional metric is considered in the Conclusions, Sec.~\ref{CONCLUSIONS}, where it is argued that consistency requires the $(\vec{x},\vec{x}{\, '})$-plane to form a $(d+d)$-dimensional Minkowski space. 

\subsubsection{Position measurement}

To introduce position measurement in the smeared-space model, let us consider the wave function of $\ket{\Psi}$, denoted as
\begin{eqnarray} \label{Psi_g} 
\Psi(\vec{x},\vec{x}{\, '}) := g(\vec{x}{\, '}-\vec{x})\psi(\vec{x}) \, , 
\end{eqnarray} 
and provide its interpretation. 
We recall that the wave function $\psi(\vec{x})$ represents the probability amplitude for obtaining the result `$\vec{x}$' from a position measurement in canonical QM, in which the background space is fixed and classical.

In our model, the smearing function $g(\vec{x}{\, '}-\vec{x})$ is interpreted as the probability amplitude for the transition $\vec{x} \mapsto \vec{x}{\, '}$.  
The wave function $\Psi(\vec{x},\vec{x}{\, '})$ therefore represents the probability amplitude for obtaining the result `$\vec{x}{\, '}$' from a position measurement in smeared-space, if the particle were to be found at the point $\vec{x}$ in the (hypothetical) fixed background.
Since an observed value `$\vec{x}{\, '}$' does not determine which classical point(s) underwent the transition $\vec{x} \mapsto \vec{x}{\, '}$ in the smeared geometry, we must sum over all possibilities by integrating the joint probability density 
$|\Psi(\vec{x},\vec{x}{\, '})|^2 $ over ${\rm d}^d\vec{x}$, yielding:
\begin{eqnarray} \label{EQ_XPRIMEDENSITY}
\frac{{\rm d}^{d}P(\vec{x}{\, '} | \Psi)}{{\rm d}\vec{x}{\, '}^{d}} &=& \int |\Psi(\vec{x},\vec{x}{\, '})|^2 {\rm d}^d\vec{x} \nonumber \\
& = & \Tr\left(\proj{\Psi} (\hat{\openone} \otimes \proj{\vec{x}{\, '}})\right)
\nonumber \\
&=& (|g|^2 * |\psi|^2)(\vec{x}{\, '}) \, .
\end{eqnarray}
This represents the generalised Born rule for position measurements in the smeared-space model. We note that in the unsmeared limit (\ref{unsmeared_limit}) it reduces to the standard Born rule of canonical QM. 

In order to give a complete description of the position measurement let us also describe the post-measurement state.
Using Eq.~(\ref{EQ_XPRIMEDENSITY}) and the definition of $\ket{\Psi}$ given in (\ref{EQ_PSIG}), an arbitrary pre-measurement state may be written as
\begin{equation} 
\ket{\Psi} = \int \sqrt{\frac{{\rm d}^{d}P(\vec{x}{\, '} | \Psi)}{{\rm d}\vec{x}{\, '}^{d}}}\ket{\psi_{\vec{x}{\, '}}} \ket{\vec{x}{\, '}} {\rm d}^{d}\vec{x}{\, '} \, ,  
\end{equation} 
where
\begin{equation} \label{post-meas_x-unprimed}
\ket{\psi_{\vec{x}{\, '}}} := \left(\frac{{\rm d}^{d}P(\vec{x}{\, '} | \Psi)}{{\rm d}\vec{x}{\, '}^{d}} \right)^{-\frac{1}{2}} \int g(\vec{x}{\, '}-\vec{x}) \psi(\vec{x}) \ket{\vec{x}} {\rm d}^d\vec{x} \, . 
\end{equation}
Hence, after measuring the value $\vec{x}{\, '} = \vec{r}_1$, the state in the fixed-background subspace of the tensor product space (corresponding to the unprimed degrees of freedom) collapses to $\ket{\psi_{\vec{r}_1}}$.
Note that this state depends on the form of the smearing function $g$, and is parameterised by $\vec{r}_1$.
We then obtain the full post-measurement state in the smeared space by applying the map (\ref{EQ_POINTMAP}) to $\ket{\psi_{\vec{r}_1}}$.
Written explicitly, this gives
\begin{equation} \label{post-meas_x}
\ket{\Psi_{\vec{r}_1}} := \int \int \frac{g(\vec{r}_1-\vec{x})g(\vec{x}{\, '}-\vec{x})\psi(\vec{x})}{\sqrt{(|g|^2 * |\psi|^2)(\vec{r}_1)}} \ket{\vec{x}}\ket{\vec{x}{\, '}} {\rm d}^d\vec{x} {\rm d}^d\vec{x}{\, '} \, .
\end{equation}
An implication of this prescription is that the system retains memory about past measurement outcomes.  
Successive post-measurement states $\ket{\Psi_{\vec{r}_1 \dots \vec{r}_n}}$ may be constructed in like manner for all $n \in \mathbb{N}$, i.e. for a sequence of position measurements with outcomes $(\vec{r}_1, \dots, \vec{r}_n)$.
Thus, the integrand of the post-measurement state, after $n$ measurements, depends on the original state $\ket{\Psi}$ and the product of $n$ additional smearing functions, each centred on one of the $n$ measurement outcomes.

Equation (\ref{EQ_XPRIMEDENSITY}) also suggests a natural definition of a generalised position observable, which may be used as a convenient tool to calculate the statistics of position measurements in the smeared-space background.
The generalised position observable $\hat{X}^{i}$, providing the $i$th component of the position vector, which acts on the smeared-state $\ket{\Psi}$, is:
\begin{eqnarray} \label{X_operator}
\hat{X}^{i} := \int x'^{i} {\rm d}^d\hat{\mathcal{P}}_{\vec{x}{\, '}} = \hat{\openone} \otimes \hat{x}'^{i} \, ,
\end{eqnarray}
where ${\rm d}^d\hat{\mathcal{P}}_{\vec{x}{\, '}} := \hat{\openone} \otimes \proj{\vec{x}{\, '}} {\rm d}^d\vec{x}{\, '}$.
It follows that
\begin{eqnarray} \label{X^n_operator}
(\hat{X}^{i})^{n} = \int (x'^{i})^{n} {\rm d}^d\hat{\mathcal{P}}_{\vec{x}{\, '}} = \hat{\openone} \otimes (\hat{x}'^{i})^{n} \, ,
\end{eqnarray}
for $n \in \mathbb{N}$, via successive applications of $\hat{X}^{i}$.
Since $\hat{X}^{i}$ is Hermitian, Eq. $(\ref{X^n_operator})$ also holds true for all $n \in \mathbb{R}$ by the spectral theorem \cite{Ish95}.

It is straightforward to verify that $\braket{\Psi |(\hat{X}^{i})^{n}|\Psi}$ gives the $n$th moment of the probability density (\ref{EQ_XPRIMEDENSITY}), for position measurements in the $i^{\rm th}$ coordinate direction. 
The associated variance is:
\begin{eqnarray} \label{X_g_uncertainty}
(\Delta_\Psi X^{i})^2 &=& \braket{\Psi |(\hat{X}^{i})^{2}|\Psi} - \braket{\Psi|\hat{X}^{i}|\Psi}^2
\nonumber\\
&=& (\Delta_\psi x'^{i})^2 + (\sigma_g^i)^2,
\end{eqnarray}
where $(\Delta_\psi x^{i})^2 = \langle \psi | (\hat x^{i})^2 | \psi \rangle - \langle \psi | \hat x^{i} | \psi \rangle^2$ is the position variance of the wave function $\psi(\vec{x})$ in the $i^{\rm th}$ coordinate direction of the fixed background of canonical QM.
We stress that the latter is just a convenient mathematical tool.
The quantum mechanical uncertainty of the smeared-space system, $\Delta_\Psi X^{i}$, may then be formally identified with the standard deviation of the probability distribution (\ref{EQ_XPRIMEDENSITY}).
Hence, as claimed, in the smeared-space model there exists a minimum position uncertainty in each spatial dimension, given by $\sigma_g^i$.

\subsubsection{Momentum measurement} \label{MOMENTUM}
 
In the fixed-background theory (i.e., canonical QM) an arbitrary quantum state $\ket{\psi}$ can be represented as an expansion in either the position or the momentum basis, giving the usual Fourier relations:
\begin{eqnarray}
	\psi(\vec{x}) & = & \frac{1}{\sqrt{2 \pi \hbar}} \
		\int \tilde{\psi}_{\hbar} (\vec{p}) \
		e^{\frac{i}{\hbar} \vec{p}.\vec{x}} \, {\rm d}^d\vec{p}  \, ,  \\  \label{psi(x)}
	\tilde \psi_{\hbar} (\vec{p}) & = &
	 \frac{1}{\sqrt{2 \pi \hbar}} \
		\int \psi (\vec{x}) \
		e^{-\frac{i}{\hbar} \vec{p}.\vec{x}} \, {\rm d}^d\vec{x}  \, . \\ \nonumber \label{psi(p)}  
\end{eqnarray}
The scale of the Fourier transforms is set by $\hbar$, which is equivalent to assuming the standard expression for the position space representation of a momentum eigenstate, 
\begin{equation} \label{EQ_DEBROGLIE}
\langle \vec{x} | \vec{p} \rangle =  \frac{1}{\sqrt{2\pi \hbar}}\
	e^{\frac{i}{\hbar} \vec{p}.\vec{x}} \, .
\end{equation}
This, in turn, follows directly from the de Broglie relation for momentum, $\vec{p} = \hbar \vec{k}$, which applies only to the wave functions of particles propagating on a classical background geometry.

We now consider physical arguments for the existence of a minimum momentum spread.
We then show that, within our formalism, the presence of minimum resolvable position- and momentum-scales implies a modification (though minute in magnitude) of the standard de Broglie relation (\ref{EQ_DEBROGLIE}).
However, crucially, our proposed modification does not significantly alter the form of the position-momentum commutator. 
Specifically, the modified observables, $\hat{X}^{i}$ and $\hat{P}_{j}$, which satisfy the new de Broglie relation, also satisfy a rescaled Heisenberg algebra. 
In this, $\hbar \mapsto \hbar + \beta$, with minute $\beta$, but the commutator remains proportional to the identity matrix. 
This is a key feature of our formalism, which permits us to recover GUP and EUP phenomenology without violating the equivalence principle \cite{Tawfik:2015rva,Tawfik:2014zca}. 
This point is discussed further in the Conclusions, Sec.~\ref{CONCLUSIONS}.

We begin with the observed vacuum energy density,
\begin{equation} \label{DE_density}
	\rho_{\Lambda} := \frac{\Lambda c^2}{8 \pi G} \simeq \
	10^{-30} \, \text{g} \cdot \text{cm}^{-3} \, ,
\end{equation}
where $\Lambda \simeq 10^{-56} \, \text{cm}^{-2}$ is the cosmological constant \cite{Hobson:2006se}.
In $(3+1)$-dimensional general relativity, this density gives rise to a maximum horizon distance of order
\begin{equation}  \label{EQ_DESITTERLENGTH}
	l_{\rm dS} := \sqrt{\frac{3}{\Lambda}} \simeq 10^{28} \, \text{cm} \, ,
\end{equation}
for any observer \cite{Spradlin:2001pw}.
This length is known as the de Sitter length and is comparable to the present day radius of the universe
\footnote{In some of the literature, $\sqrt{3/\Lambda}$ is also referred to as the Wesson length (denoted $l_{\rm W}$) after the pioneering work \cite{Wesson:2003qn}.}.
Hence, the maximum position uncertainty for any particle in a classical background geometry, with minimum energy density $\rho_{\Lambda}$, is $(\Delta_{\psi} x)_{\max} \simeq l_{\rm dS}$ (in any coordinate direction).  
By the HUP, the corresponding minimum momentum uncertainty is of order $(\Delta_{\psi} p)_{\min} \simeq \hbar / l_{\rm dS} \simeq m_{\rm dS} c$, where
\begin{equation} \label{EQ_DESITTERMASS}
m_{\rm dS} := \frac{\hbar}{c} \sqrt{\frac{\Lambda}{3}} \simeq 10^{-66} \, \text{g} 
\end{equation}
is the de Sitter mass.
Hence, we fix the smearing-scale for three-dimensional momentum space to be of the order of the de Sitter momentum, $\sim m_{\rm dS}c$, where $m_{\rm dS}$ is given by Eq. (\ref{EQ_DESITTERMASS}). 
In $D = d+1$ space-time dimensions, the dark energy density is given by
\begin{equation} \label{DE_density_HD}
\rho_{\Lambda_{D}} := \frac{\Lambda_{D}c^2}{2d \, \Omega_{d}G_{D}} \, , 
\end{equation}
where $G_{D}$ is the $D$-dimensional Newton's constant and $\Omega_{d} = \pi^{d/2}/\Gamma(d/2+1)$ is the volume of the unit $d$-sphere. 
In the presence of $d-3$ compactified (or warped) dimensions, 
$G_{D} \simeq GV_{d-3}$  on length-scales greater than $(V_{d-3})^{\frac{1}{d-3}}$, where $V_{d-3}$ is the volume of the internal space \cite{Maartens:2010ar}. 
Hence, $\rho_{\Lambda_{D}} \simeq \rho_{\Lambda}/V_{d-3}$ and $\Lambda_{D}/d \simeq (\Lambda/3)(\Omega_{d}/\Omega_3)$.
The $D$-dimensional de Sitter length- and mass-scales are then:
\begin{eqnarray}  \label{EQ_DESITTERSCALES_HD}
l_{\rm dS} := \sqrt{\frac{d}{\Lambda_D}} \, , \quad m_{\rm dS} := \frac{\hbar}{c} \sqrt{\frac{\Lambda_{D}}{d}} \, . 
\end{eqnarray}
In the following analysis, $l_{\rm dS}$ and $m_{\rm dS}$ are used to denote the de Sitter scales 
in an arbitrary (unspecified) number of dimensions, unless otherwise stated.

Next, by analogy with our description of the smearing of position space, we introduce the amplitude $\tilde g_\beta(\vec{p}{ \, '}- \vec{p})$, whose squared modulus gives the probability that a point $\vec{p}$ in classical momentum space undergoes the transition $\vec{p} \mapsto \vec{p}{ \, '}$. 
(The meaning of the index $\beta$ will be made clear soon.)
Hence, we impose that the momentum space representation of the smeared-space wave function is analogous to its position space representation, i.e.,
\begin{equation} \label{EQ_TILDEPSI}
\tilde \Psi (\vec{p},\vec{p}{ \, '}) := \tilde g_\beta(\vec{p}{ \, '} - \vec{p}) \, \tilde \psi_{\hbar}(\vec{p}) \, .
\end{equation}
We then choose a basis $\ket{\vec{p} \, \vec{p}{ \, '}}$ in the tensor product Hilbert space which ensures that Eq. (\ref{EQ_TILDEPSI}) holds. 

Consider the following map from a state $\ket{\vec{p}}$ in the classical background to a state in the smeared-space:
\begin{eqnarray} \label{EQ_MOMENTUM_MAP}
\ket{\vec{p}} \mapsto \int \tilde{g}_{\beta}(\vec{p}{ \, '}-\vec{p})\ket{\vec{p} \, \vec{p}{ \, '}}{\rm d}^d\vec{p}{ \, '},
\end{eqnarray}
where $\ket{\vec{p} \, \vec{p}{ \, '}}$ denotes the basis vector labeled by $\vec{p}$ and $\vec{p}{ \, '}$, which need not be a simple tensor product. 
(We stress this by not writing a comma in between $\vec{p}$ and $\vec{p}{ \, '}$, in contradistinction to the position space basis, $|\vec{x},\vec{x}'\rangle := |\vec{x}\rangle |\vec{x}'\rangle$.)
Applying the map (\ref{EQ_MOMENTUM_MAP}) to a state $\ket{\psi}$ in a fixed momentum space background, i.e. $\ket{\psi} = \int \tilde \psi_{\hbar}(\vec{p}) \ket{\vec{p}} {\rm d}^d\vec{p}$, gives
\begin{eqnarray} \label{EQ_PSIG*}
\ket{\Psi} := \int\int \tilde{g}_{\beta}(\vec{p}{ \, '}-\vec{p})\tilde{\psi}_{\hbar}(\vec{p})\ket{\vec{p} \, \vec{p}{ \, '}} {\rm d}^d\vec{p} {\rm d}^d\vec{p}{ \, '} \, ,
\end{eqnarray}
where ${\rm d}^d\vec{p} = \sqrt{\det \tilde{g}_{ij}(p)}{\rm d}^dp$, ${\rm d}^d\vec{p}{\, '} = \sqrt{\det \tilde{g}'_{ij}(p')}{\rm d}^dp'$. 
Here, $\tilde{g}_{ij}(p)$ and $\tilde{g}'_{ij}(p')$ denote the metrics on the sub-spaces defined by $\vec{p}{\, '} = {\rm const.}$ and $\vec{p} = {\rm const.}$, respectively. 

Expansion in the basis $\ket{\vec{p} \, \vec{p}{ \, '}}$ then forms the momentum space representation for all states in the smeared-space model, i.e., $\tilde \Psi (\vec{p},\vec{p}{ \, '}) = \langle \vec{p} \, \vec{p}{ \, '} | \Psi \rangle$.
We obtain Eq.~(\ref{EQ_TILDEPSI}) by setting
\begin{eqnarray} \label{EQ_PIPIPRIME}
| \vec{p} \, \vec{p}{ \, '} \rangle &:=& \frac{1}{2 \pi \sqrt{\hbar \beta}} 
        \nonumber\\
	&\times& \int \! \int  \
	e^{\frac{i}{\hbar} \vec{p}.\vec{x}} \, e^{\frac{i}{\beta}(\vec{p}{\, '}-\vec{p}).(\vec{x}{\, '}-\vec{x})} \
	|\vec{x} \rangle |\vec{x}' \rangle \, {\rm d}^d\vec{x} {\rm d}^d\vec{x}{\, '} \, ,
\nonumber\\
\end{eqnarray}
where the position and momentum smearing functions are related by the Fourier transforms at scale $\beta$:
\begin{equation}
g(\vec{x}{\, '} - \vec{x}) := \frac{1}{\sqrt{2 \pi \beta}} \int \tilde g_\beta(\vec{p}{ \, '} - \vec{p}) \, e^{\frac{i}{\beta}(\vec{p}{ \, '}-\vec{p}).(\vec{x}{\, '}-\vec{x})} {\rm d}^d\vec{p}{ \, '} \, ,
\end{equation}
\begin{equation}
\tilde g_\beta(\vec{p}{ \, '} - \vec{p}) := \frac{1}{\sqrt{2 \pi \beta}} \int g(\vec{x}{\, '}-\vec{x}) \, e^{- \frac{i}{\beta}(\vec{p}{ \, '}-\vec{p}).(\vec{x}{\, '}-\vec{x})} {\rm d}^d\vec{x}{\, '} \, .
\end{equation}

The fact that $\tilde{g}_{\beta}(\vec{p}{ \, '}-\vec{p})$ is the Fourier transform of $g(\vec{x}{\, '}-\vec{x})$, transformed at the scale $\beta$ rather than $\hbar$, implies a kind of wave-point duality, analogous to the wave-particle duality of canonical quantum mechanics.
In the canonical formalism, the conjugate variable to position, $\vec{x}$, is the wave vector, $\vec{k}$, which gives rise to the uncertainty principle $\Delta_{\psi}x^{i} \, \Delta_{\psi}k_{j} \geq (1/2) \ \delta^{i}{}_{j}$.
The wave vector is related to the `particle' momentum by the scale factor, $\hbar$, through the de Broglie relation $\vec{p} = \hbar \vec{k}$.
This yields the HUP and the scale for the transformations between $\psi(\vec{x})$ and $\tilde{\psi}_{\hbar}(\vec{p})$. 
Here, we use the subscript $\hbar$ to emphasise this point.

Similarly, in the smeared-space theory, the conjugate variable to $(\vec{x}{\, '} - \vec{x})$ is $(\vec{k}' - \vec{k})$, which is now related to $\vec{p}{ \, '}-\vec{p}$ by the scale $\beta$, i.e., such that $\vec{p}{ \, '}-\vec{p} = \beta(\vec{k}'-\vec{k})$. 
Here, $\vec{p}{ \, '}-\vec{p}$ refers to the momentum associated with a smeared spatial `point', rather than a point-particle on a fixed background. 
However, in a given classical background, $\vec{p}$ retains its standard interpretation as the momentum of a particle, and we assume that the standard de Broglie relation $\vec{p}=\hbar \vec{k}$ holds, together with the relation above. 
We then have:
\begin{eqnarray} \label{mod_dB}
\vec{p}{ \, '}= \hbar \vec{k} +\beta (\vec{k}' -\vec{k}) \, .
\end{eqnarray}
This may be regarded as the modified de Broglie relation for particles on the smeared-space background.
Equation (\ref{mod_dB}) follows directly from the relation
\begin{equation} \label{smeared-p-eigenstate}
\langle \vec{x} | \langle \vec{x}{\, '} | \vec{p} \, \vec{p}{ \, '} \rangle = \frac{1}{2 \pi \sqrt{\hbar \beta}} \
	e^{\frac{i}{\hbar} \vec{p}.\vec{x}} \, e^{\frac{i}{\beta}(\vec{p}{ \, '}-\vec{p}).(\vec{x}{\, '}-\vec{x})} \, ,
\end{equation}
which is the smeared-space generalisation of Eq. (\ref{EQ_DEBROGLIE}).

It follows from the general properties of the Fourier transform~\cite{pinsky2008introduction} that
\begin{equation} \label{EQ_BETA_SCALE}
\Delta_g x'^{i} \, \Delta_g p'_{j} \geq \frac{\beta}{2} \, \delta^{i}{}_{j} \, .
\end{equation}
This may be regarded as the uncertainty principle for spatial `points', as opposed to point-particles on a classical spatial background. 
Thus, choosing the squares of the smearing functions $|g(\vec{x}{\, '} - \vec{x})|^2$ and $|\tilde{g}_{\beta}(\vec{p}{\, '} - \vec{p})|^2$ to be normalised Gaussian distributions, with standard deviations $\Delta_g x'^{i} = \sigma_{g}^{i}$ and $\Delta_g p'_{j} = \tilde{\sigma}_{g j}$ for all $i,j \in \left\{1,2, \dots ,d\right\}$, the inequality in (\ref{EQ_BETA_SCALE}) is saturated, yielding the definition of the transformation scale $\beta$: 
\begin{equation} \label{beta-1}
\sigma_{g}^{i} \, \tilde{\sigma}_{gj} =: \frac{\beta}{2} \, \delta^{i}{}_{j} \, ,
\end{equation}
or, equivalently, $\beta := (2/d)\sigma_{g}^{i} \, \tilde{\sigma}_{gi}$.

We now fix exact values of the parameters $\sigma_{g}^{i}$ and $\tilde{\sigma}_{gi}$, and hence the scale $\beta$, from physical considerations. 
In $(3+1)$ space-time dimensions, equating the reduced Compton wavelength $\lambda_{\rm C} = \hbar/(mc)$ and Schwarzschild radius $r_{\rm S} = 2Gm/c^2$, of a mass $m$, gives $\lambda_{\rm C}=r_{\rm S} = \sqrt{2}l_{\rm Pl}$, $m = m_{\rm max} := (1/\sqrt{2})m_{\rm Pl}$, where 
\begin{eqnarray} \label{PLANCKLENGTH_3D}
l_{\rm Pl} := \sqrt{\hbar G/c^3} \simeq 10^{-33} \, {\rm cm} \, , 
\end{eqnarray}
and
\begin{eqnarray} \label{PLANCKMASS_3D}
m_{\rm Pl} := \sqrt{\hbar c/G} \simeq 10^{-5} \, {\rm g} \, , 
\end{eqnarray}
are the Planck length- and mass-scales, respectively.
This marks the boundary on the mass-radius diagram between the quantum (particle) and gravitational (black hole) domains \cite{Carr:2014mya}.
Thus, we take the minimum position uncertainty to be $\sqrt{2}l_{\rm Pl}$, where $l_{\rm Pl}$ is given by Eq. (\ref{PLANCKLENGTH_3D}).

In $D = d+1$ space-time dimensions, for arbitrary $d$, the Schwarzschild radius is $r_{\rm S} = (2G_{D}m/c^2)^{\frac{1}{d-1}}$ \cite{Horowitz:2012nnc}. 
The intersection of the Compton and Schwarzschild lines is then given by $\lambda_{\rm C}=r_{\rm S} = 2^{\frac{1}{d-1}}l_{\rm Pl}$, $m = m_{\rm max} := 2^{-\frac{1}{d-1}}m_{\rm Pl}$, where
\begin{eqnarray} \label{PLANCKSCALES_HD}
l_{\rm Pl} := (\hbar G_{D}/c^3)^{\frac{1}{d-1}} \, , \quad m_{\rm Pl} := (\hbar^{d-2} c^{4-d}/G_{D})^{\frac{1}{d-1}} \, , 
\end{eqnarray}
are the $D$-dimensional Planck length- and mass-scales, respectively \cite{Maartens:2010ar}.
In the following analysis, $l_{\rm Pl}$ and $m_{\rm Pl}$ are used to denote the Planck 
scales in an arbitrary (unspecified) number of dimensions, unless otherwise stated.

As discussed above Eq. \eqref{EQ_DESITTERMASS}, taking the de Sitter scale as the maximum position uncertainty, the HUP implies a corresponding minimum momentum uncertainty. 
The smearing-scale for momentum space is therefore taken to be one-half the de Sitter momentum, regardless of the dimensionality of space-time.
Hence, we define
\begin{equation} \label{sigma_g}
\sigma_g^{i} := 2^{\frac{1}{d-1}}l_{\rm Pl} \, , \quad \tilde{\sigma}_{gi} := \frac{1}{2}m_{\rm dS}c \, ,
\end{equation} 
for all linear coordinate directions, yielding 
\begin{equation} \label{beta-2_HD}
\beta := 2^{\frac{d+1}{2(d-1)}} \hbar\sqrt{\frac{\rho_{\Lambda_D}}{\rho_{\rm Pl}}} \, , 
\end{equation}
where  $\rho_{\rm Pl} := m_{\rm Pl}/(\Omega_{d}l_{\rm Pl}^d)$ is the $D$-dimensional Planck density.
In $(3+1)$ dimensions (our observable universe) this gives:
\begin{equation} \label{beta-2}
\beta := 2 \hbar\sqrt{\frac{\rho_{\rm \Lambda}}{\rho_{\rm Pl}}} \simeq \hbar \times 10^{-61} \, ,
\end{equation}
where $\rho_{\rm Pl} = (3/4\pi)m_{\rm Pl}/l_{\rm Pl}^3 \simeq 10^{93} \, {\rm g  \cdot cm^{-3}}$.

Note that the wave-point duality implied by Eq. (\ref{EQ_BETA_SCALE}) requires a finite nonzero value of $\beta$. 
This, in turn, requires finite nonzero values of both $\sigma_g^{i}$ and $\tilde \sigma_{gi}$.
In principle, finite $\beta$ could also be obtained in the limit $\tilde \sigma_{gi} \rightarrow 0$, $\sigma_g^{i} \rightarrow \infty$ or $\tilde \sigma_{gi} \rightarrow \infty$, $\sigma_g^{i} \rightarrow 0$.  
However, the former case gives rise to an unnormalisable $g(\vec{x}{\, '}-\vec{x})$, where each point is spread uniformly over all physical space.
Similarly, the latter gives rise to an unnormalisable $\tilde{g}_{\beta}(\vec{p}{ \, '}-\vec{p})$.
In other words, it is impossible, within our formalism, to self-consistently smear only position or momentum space.  
The physical implications of this are discussed in Sec. \ref{Cosmology}.

In full analogy to the case of position measurement, the probability density associated with the observed momentum $\vec{p}{ \, '}$ is given by:
\begin{eqnarray} \label{EQ_PPRIMEDENSITY}
\frac{{\rm d}^dP(\vec{p}{ \, '} | \Psi)}{{\rm d}\vec{p}{ \, '}^d} &=& \int |\tilde \Psi(\vec{p},\vec{p}{ \, '})|^2 {\rm d}^d\vec{p} \nonumber \\
&=& (|\tilde g_\beta|^2 * |\tilde \psi_{\hbar}|^2)(\vec{p}{ \, '}) \, .
\end{eqnarray}
One then verifies that the moments of this distribution are given by the brackets $\braket{\Psi|(\hat{P}_{j})^{n}|\Psi}$, where the generalised momentum operator is defined as
\begin{eqnarray} \label{P_operator} 
\hat{P}_{j} &:=& 
\int \int p'_{j} |\vec{p} \, \vec{p}{ \, '}\rangle \langle \vec{p} \, \vec{p}{ \, '}| {\rm d}^d\vec{p} {\rm d}^d\vec{p}{ \, '} \, .
\end{eqnarray}
The uncertainty of smeared-space momentum measurements in the $j^{\rm th}$ coordinate direction is, therefore:
\begin{eqnarray} \label{P_g_uncertainty}
(\Delta_\Psi P_{j})^2 &=& \braket{\Psi|(\hat{P}_{j})^{2}|\Psi} - \braket{\Psi|\hat{P}_{j}|\Psi}^2
\nonumber\\
&=& (\Delta_\psi p'_{j})^2 + \tilde{\sigma}_{g j}^2 \, ,
\end{eqnarray}
where $(\Delta_\psi p_{j})^2 = \langle \psi | (\hat p_{j})^2 | \psi \rangle - \langle \psi | \hat p_{j} | \psi \rangle^2$ is the momentum variance of $\tilde{\psi}_{\hbar}(\vec{p})$ in the $j^{\rm th}$ coordinate direction of the fixed-background theory.

To complete the description of momentum measurement, let us explain how to obtain the post-measurement state.
From Eq.~(\ref{EQ_PPRIMEDENSITY}), and using the fact that the states $\ket{\vec{p} \, \vec{p}{\, '}}$ are orthogonal, $\langle \vec{q} \, \vec{q}{\, '} | \vec{p} \, \vec{p}{\, '} \rangle = \delta^d(\vec{q}-\vec{p}) \delta^d(\vec{q}{\, '}-\vec{p}{\, '})$, any initial (pre-measurement) state can be written as
\begin{eqnarray} \label{EQ_MOM_PM}
\ket{\Psi} &=& \int \sqrt{\frac{{\rm d}^dP(\vec{p}{\, '}|\Psi)}{{\rm d}\vec{p}{\, '}^d}} 
\nonumber\\
&\times& \left(\int \frac{\tilde{g}_{\beta}(\vec{p}{\, '}-\vec{p})\tilde{\psi}_{\hbar}(\vec{p})}{\sqrt{(|\tilde g_\beta|^2 * |\tilde \psi_\hbar|^2)(\vec{p}{\, '})}} \ket{\vec{p} \, \vec{p}{\, '}} {\rm d}^d\vec{p} \right) {\rm d}^d\vec{p}{\, '} \, ,
\end{eqnarray}
where in the bracket we indicate the state labelled by a fixed value of $\vec{p}{\, '}$.
In contrast to the case of smeared-space position measurements, where a fixed value of $\vec{x}{\, '}$ indicates a definite fixed-background state (labelled by $\vec{x}{\, '}$), this is no longer the case in smeared momentum space.
Since the basis vectors $\ket{\vec{p} \, \vec{p}{\, '}}$ are entangled, it is not possible to identify a definite state in the fixed-background theory, labelled by $\vec{p}{\, '}$.

However, it is not necessary to identify a definite fixed-background state in order to obtain the final post-measurement state. 
Instead of (\ref{EQ_MOMENTUM_MAP}), one can define the following map that acts on the basis $\ket{\vec{p} \, \vec{p}{\, '}}$, spanning both the primed and unprimed subsystems:
\begin{equation} \label{EQ_POINTMAP*} 
\ket{\vec{p} \, \vec{s}{\, '}} \mapsto \int \tilde{g}_{\beta}(\vec{p}{\, '}-\vec{p})\ket{\vec{p} \, \vec{p}{\, '}} {\rm d}^d\vec{p}{\, '} \, . 
\end{equation}
The final post-measurement state is obtained by applying this map to the state inside the bracket in Eq.~(\ref{EQ_MOM_PM}). 
Assuming that the value $\vec{s}_1$ was obtained in the momentum measurement, the resulting post-measurement state may be written explicitly as:
\begin{equation} \label{post-meas_p}
\ket{\Psi_{\vec{s}_1}} := \int \int \frac{ \tilde{g}_{\beta}(\vec{s}_1-\vec{p}) \tilde{g}_{\beta}(\vec{p}{\, '}-\vec{p}) \tilde{\psi}_{\hbar}(\vec{p})}{\sqrt{(|\tilde g_\beta|^2 * |\tilde \psi_\hbar|^2)(\vec{s}_1)}} \ket{\vec{p} \, \vec{p}{\, '}} {\rm d}^d\vec{p} {\rm d}^d\vec{p}{\, '} \, .
\end{equation}
This is analogous to Eq. (\ref{post-meas_x}). 
Again, subsequent post-measurement states $\ket{\Psi_{\vec{s}_1 \dots \vec{s}_n}}$, corresponding to sequential momentum measurements in the smeared-space model, 
may be constructed in like manner and depend on the values measured before.

\subsection{The smeared-operator picture} \label{smeared_ops}

Up to now, we have described the effect of smearing the background space on which quantum particles propagate by modifying the canonical quantum wave function, mapping $\ket{\psi} \mapsto \ket{\Psi}$.
We now briefly discuss an alternative approach, in which the quantum states associated with particles remain unsmeared, but in which the observables that act on them are smeared.
Both formulations give rise to identical predictions for the generalised position and momentum uncertainties and, in this sense, may be thought of as analogous to the Schr{\" o}dinger and Heisenberg pictures of canonical quantum theory.

\subsubsection{Smeared operators}

In order to introduce the smeared operators, let us recall the fundamental map modeling the smearing of position space, Eq. (\ref{EQ_POINTMAP}).
We introduce the smearing operator $\hat{S}$, such that
\begin{equation}
\hat S \ket{\vec{x}} := \ket{\vec{x}} \otimes \ket{g_{\vec{x}}} \, .
\end{equation}
Written explicitly, it has the following representation in the position-space basis:
\begin{eqnarray} \label{S_g}
\hat{S} &=& \int \int g(\vec{x}{\, '} - \vec{x}) (\ket{\vec{x}}\otimes \ket{\vec{x}{\, '}}) \bra{\vec{x}} {\rm d}^d\vec{x}{\rm d}^d\vec{x}{\, '} \, .
\end{eqnarray}

With this definition, an arbitrary state in smeared space is given by 
\begin{equation}
\ket{\Psi} = \hat S \ket{\psi} \, .
\end{equation}
Hence, the statistical predictions of our formalism can also be obtained as:
\begin{equation}
\langle \Psi | (\hat X^{i})^{n} | \Psi \rangle = \langle \psi | \hat S^\dagger (\hat X^{i})^{n} \hat S |\psi \rangle  \, ,
\end{equation} 
i.e., by using the Hermitian operator $(\hat{X}^{i})^{(n)}_S := \hat{S}^\dagger (\hat X^{i})^{n} \hat{S}$, together with the fixed-background state $\ket{\psi}$.
More explicitly, the Hermitian operator reads:
\begin{eqnarray} \label{X_g^(n)}
(\hat{X}^{i})^{(n)}_S &=& \int \int (x'^{i})^{n} |g(\vec{x}{\, '} - \vec{x}) |^2 \ket{\vec{x}}\bra{\vec{x}} {\rm d}^d\vec{x}{\rm d}^d\vec{x}{\, '}  
\nonumber\\
&=& \int \braket{(x'^{i})^{n}}_{g} \ket{\vec{x}}\bra{\vec{x}} {\rm d}^d\vec{x} \, . 
\end{eqnarray}
Note that, here, the probability density $|g(\vec{x}{\, '}-\vec{x})|^2$ is raised to only the first power in the integrand.
Operationally, this reflects the fact that one measures the smeared-position observable, which yields the nominal value `$x'^{i}$', which is then raised to the required power $n$ in order to generate the statistics of the system. 
We emphasise that one must distinguish between these operators and those representing genuine repeated measurements.
We shall now describe such sequential smeared measurements.

\subsubsection{Sequential measurements}

A property of measurements which is crucial for the consideration of their sequences is that, in general, they modify the state of the measured object.
We now show that, in the smeared-operator picture, the required state-update procedure is particularly simple.

As shown above, the pre-measurement state of the smeared-space quantum system $\ket{\Psi}$ may be obtained by applying the smearing operator $\hat S$ to an arbitrary fixed-background state, $\ket{\psi}$.
In the smeared-state picture of position measurement, the generalised projection associated with the outcome $x'^{i}$ is then ${\rm d}^d\hat{\mathcal{P}}_{\vec{x}{\, '}} = \hat{\openone} \otimes \proj{\vec{x}{\, '}} {\rm d}^d\vec{x}{\, '}$.
This acts on the state $\ket{\Psi}$. 
We now note that the statistics of this projective measurement may also be obtained from another set of measurement operators, defined as $\hat M_{\vec{x}{\, '}} := \hat{\openone} \otimes \bra{\vec{x}{\, '}}$,
i.e., such that $\hat M_{\vec{x}{\, '}} (|\vec{a}\rangle \otimes |\vec{a}{\, '}\rangle) = \langle \vec{x}{\, '} | \vec{a}{\, '} \rangle \, |\vec{a}\rangle$.
(Of course, $\hat M_{\vec{x}{\, '}}^{\dagger} | \psi \rangle = | \psi \rangle \otimes |\vec{x}{\, '}\rangle$ 
and, by this definition, the generalised projectors in the smeared-state picture may be written as ${\rm d}^d\hat{\mathcal{P}}_{\vec{x}{\, '}} = \hat M_{\vec{x}{\, '}}^\dagger \hat M_{\vec{x}{\, '}} {\rm d}^d\vec{x}{\, '}$.)

In this way, a smeared position measurement performed on the state $\ket{\Psi}$, yielding outcome $\vec{x}{\, '} = \vec{r}_1$, leaves the post-measurement state
\begin{equation}
\ket{\Psi_{\vec{r}_1}} = \frac{\hat S \, \hat M_{\vec{r}_1} \, \hat S \ket{\psi}}{\sqrt{\langle \psi | \hat S^\dagger \hat M_{\vec{r}_1}^\dagger \hat M_{\vec{r}_1} \hat S | \psi \rangle}} \, ,
\end{equation}
where $\ket{\Psi_{\vec{r}_1}}$ is given by Eq. (\ref{post-meas_x}). 
Here, the first $\hat S$ operator smears the initial state $\ket{\psi}$. 
The measurement operator $\hat M_{\vec{r}_1}$ (together with the normalisation factor) then collapses the fixed-background part of the tensor product to the post-measurement state corresponding to the value $\vec{x}{\, '} = \vec{r}_1$, see Eq.~(\ref{post-meas_x-unprimed}).
Finally, the second $\hat S$ operator re-smears the post-measurement state of the fixed background.
A sequence of such measurements, which generates a sequence of outcomes $(\vec{r}_1,\dots, \vec{r}_n)$, produces the final post-measurement state that one obtains by applying the sequence of smeared operators:
\begin{equation}
\ket{\Psi_{\vec{r}_1 \dots \vec{r}_n}} = \frac{ \hat S \, \hat M_{\vec{r}_n} \, \hat S \dots \hat M_{\vec{r}_2} \, \hat S \hat M_{\vec{r}_1} \, \hat S \ket{\psi}}
{\sqrt{\langle \psi | \hat S^\dagger \hat M_{\vec{r}_1}^\dagger \dots \hat S^\dagger \hat M_{\vec{r}_n}^\dagger \hat M_{\vec{r}_n} \hat S \dots \hat M_{\vec{r}_1} \hat S | \psi \rangle}} \, .
\end{equation}
As mentioned, the repeated measurements do not leave the system invariant and it possesses memory of past measurement outcomes.
Similar considerations hold for the family of smeared-momentum operators $(\hat{P}_{j})^{(n)}_S$, which may be defined in full analogy to the case of position measurement.

We may also define the family of smeared Hermitian operators associated with a general Hermitian observable $\hat{O}$ in the smeared-state picture, i.e., $\hat{O}^{(n)}_S := \hat{S}^{\dagger}\hat{O}^{n}\hat{S}$. 
In particular, we note that a general commutator $[\hat{O}_1,\hat{O}_2]$, where $\hat{O}_1$, $\hat{O}_2$ act on the smeared-state $\ket{\Psi}$, is mapped to $[\hat{O}_1,\hat{O}_2]_{S} := \hat{S}^{\dagger}[\hat{O}_1,\hat{O}_2]\hat{S}$, where $[\hat{O}_1,\hat{O}_2]_{S}$ acts on $\ket{\psi}$.
This may be rewritten as $[\hat{O}_1,\hat{O}_2]_{S} = \hat{\mathcal{O}}_{1S}^{\dagger}\hat{\mathcal{O}}_{2S} - \hat{\mathcal{O}}_{2S}^\dagger \hat{\mathcal{O}}_{1S}$, where $\hat{\mathcal{O}}^{(n)}_S := \hat{O}^{n}\hat{S}$ and $\hat{\mathcal{O}}^{(n)\dagger}_S := \hat{S}^{\dagger}\hat{O}^{n}$.
We note that, unlike the $\hat{O}^{(n)}_S$, the operators $\hat{\mathcal{O}}^{(n)}_S$ do not posses the properties usually ascribed to quantum observables.
They have no spectral decomposition and hence no eigenvalues.
However, they can be seen as generalised measurements in the canonical theory, i.e., positive operator valued measures (POVMs) \cite{Chuang_Nielsen}.
Indeed, one verifies that the operators $\hat{\mathcal{M}}_{\vec{x}{\, '}} := \, \hat M_{\vec{x}{\, '}} \, \hat S$ form a POVM.

\subsubsection{Comments}

Both formulations of the smeared-space theory, based on smeared-states and on smeared-operators, respectively, imply a non-trivial modification of the Schr{\" o}dinger equation. 
In the former, the momentum observable $\hat{\vec{P}}$, defined via Eq. (\ref{P_operator}), acts on the smeared-state $\ket{\Psi}$, 
whereas, in the latter, $\hat{\vec{P}}$ is replaced with an appropriate smeared version, $\hat{\vec{\mathcal{P}}}_S$, that acts on the fixed-background state $\ket{\psi}$.

This is in agreement with our intuition that well-defined translations do not exist on an imprecise (smeared) background, since the position space representation of the canonical momentum operator may be identified with the generator of spatial translations up to a factor of $\hbar$ \cite{Ish95}. 
Hence, if we wish to act on the fixed-background state, the smearing must be incorporated into the operator itself. 

Equivalently, in the smeared-state picture, we may view the generalised momentum operator $\hat{\vec{P}}$ as performing precise infinitesimal translations on each geometry in a superposition of backgrounds and similar arguments apply to the momentum space representation of $\hat{\vec{X}}$. 
Though it is beyond the scope of this paper to consider these effects in detail, we here include both formulations of the smeared-space model, which may be used as a basis for further investigations. 

\subsection{Uncertainty relations} \label{SEC_UR}

We begin this section by presenting the general uncertainty relation, which follows as an immediate consequence of our previous considerations.
Next, we show that the well known uncertainty relations, the GUP (\ref{GUP-1*}) and EUP (\ref{EUP-1}), arise as limits of the individual smeared-space uncertainty relations, (\ref{X_g_uncertainty}) and (\ref{P_g_uncertainty}), respectively.
Finally, we discuss the emergence of the EGUP as a limit of the general relation.

Combining Eqs.~(\ref{X_g_uncertainty}) and (\ref{P_g_uncertainty}), we have
\begin{eqnarray} \label{X_g_P_X_g_uncertainty-2}
(\Delta_\Psi X^{i})^2 \, (\Delta_\Psi P_{j})^2 &=& \left((\Delta_\psi x'^{i})^2 + (\sigma_g^{i})^2\right) 
\nonumber\\
&\times& \left((\Delta_\psi p'_{j})^2 + (\tilde{\sigma}_{gj})^2\right) \, .
\end{eqnarray} 
The HUP then gives
\begin{eqnarray} \label{GUR_X}
(\Delta_\Psi X^{i})^2 \, (\Delta_\Psi P_{j})^2 
& \ge & (\hbar/2)^2(\delta^{i}{}_{j})^2 + (\Delta_\psi x'^{i})^2 (\tilde{\sigma}_{gj})^2 
\nonumber\\
&+& (\sigma_g^{i})^2\frac{(\hbar/2)^2}{(\Delta_\psi x'^{j})^2} + (\sigma_g^{i})^2 (\tilde{\sigma}_{gj})^2 \, ,
\nonumber\\
\end{eqnarray}
and
\begin{eqnarray} \label{GUR_P}
(\Delta_\Psi X^{i})^2 \, (\Delta_\Psi P_{j})^2 
& \ge & (\hbar/2)^2(\delta^{i}{}_{j})^2 + \frac{(\hbar /2)^2}{(\Delta_\psi p'_{i})^2} (\tilde{\sigma}_{gj})^2 
\nonumber\\
&+& (\sigma_g^{i})^2 (\Delta_\psi p'_{j})^2 + (\sigma_g^{i})^2 (\tilde{\sigma}_{gj})^2 \, .
\nonumber\\
\end{eqnarray}
Optimising the right-hand side of Eq. (\ref{GUR_X}) with respect to $\Delta_\psi x'^{i}$ yields
\begin{equation} \label{EQ_CAN_DX_OPT}
(\Delta_\psi x'^{i})_{\mathrm{opt}} := \sqrt{\frac{\hbar}{2} \frac{\sigma_g^{i}}{\tilde{\sigma}_{gi}}} \, .
\end{equation}
Note that, here, both indices inside the square root are contravariant, so that no summation is implied.
Similarly, optimising the right-hand side of Eq. (\ref{GUR_P}) with respect to $\Delta_\psi p'_{j}$ gives
\begin{equation} \label{EQ_CAN_DP_OPT}
(\Delta_\psi p'_{j})_{\mathrm{opt}} := \sqrt{\frac{\hbar}{2} \frac{\tilde{\sigma}_{gj}}{\sigma_g^{j}}} \, .
\end{equation}

In both cases, the sum of the middle two terms on the right-hand side of the relevant inequality, (\ref{GUR_X}) or (\ref{GUR_P}), is simply $(\hbar\beta/2)(\delta^{i}{}_{j})^2$, yielding
\begin{eqnarray} \label{DXDP_opt}
(\Delta_\Psi X^{i})^2 \, (\Delta_\Psi P_{j})^2 & \ge & \frac{(\hbar + \beta)^2}{4} \, (\delta^{i}{}_{j})^2 \, .
\end{eqnarray}
The same result is readily obtained by noting that the commutator of the position and momentum observables in the smeared-space formalism is:
\begin{equation} \label{[X,P]}
[\hat{X}^{i},\hat{P}_{j}] = i(\hbar + \beta)\delta^{i}{}_{j} \, {\bf\hat{\mathbb{I}}} \, ,
\end{equation}
where ${\bf\hat{\mathbb{I}}} = \hat{\openone} \otimes \hat{\openone}$ is the identity matrix on the tensor product space and $\hat{\openone}$ is the identity matrix on the Hilbert space of canonical $d$-dimensional QM.
Equation (\ref{DXDP_opt}) then follows directly from the Schr{\" o}dinger-Robertson relation (\ref{Robertson-Schrodinger-1}).
The minimum is achieved by choosing $\psi(\vec{x})$ and $\tilde{\psi}_{\hbar}(\vec{p})$ to be Gaussians with standard deviations $(\Delta_\psi x'^{i})_{\mathrm{opt}}$ (\ref{EQ_CAN_DX_OPT}) and $(\Delta_\psi p'_{i})_{\mathrm{opt}}$ (\ref{EQ_CAN_DP_OPT}), in every coordinate direction $x^{i}$, respectively. 

We emphasise that, in our formalism, the modification of the canonical uncertainty relation (HUP) does not arise from a modification of the canonical commutator of the form (\ref{modified_commutator-1}). 
Although the smeared-space theory implies a modified de Broglie relation, as in Eq. (\ref{mod_dB}), the position-momentum commutator remains proportional to the identity matrix. 
Heuristically, we can understand the non-canonical term in Eq. (\ref{[X,P]}), $i\beta \delta^{i}{}_{j} \, {\bf\hat{\mathbb{I}}} = i \, . \, 2 \sigma_g^{i} \tilde{\sigma}_{gj}  \, {\bf\hat{\mathbb{I}}}$, as arising from the modified expectation values of operators in the presence of minimum position and momentum space smearing.

Finally, we note that the position-momentum symmetry of the general relation may be quantified in terms of the optimising values (\ref{EQ_CAN_DX_OPT})-(\ref{EQ_CAN_DP_OPT}). 
More specifically, the smeared-space uncertainty relation, Eq. (\ref{X_g_P_X_g_uncertainty-2}), is invariant under the simultaneous transformations:
\begin{eqnarray} \label{DXDP_symm_transf-1}
\Delta_\psi x'^{i} &\rightarrow& \frac{(\Delta_{\psi}x'^{i})_{\rm opt}}{(\Delta_{\psi}p'_{j})_{\rm opt}} \Delta_\psi p'_{j} = \frac{\sigma_g^{i}}{\tilde{\sigma}_{gj}}\Delta_\psi p'_{j} \, , 
\nonumber\\
\Delta_\psi p'_{j} &\rightarrow& \frac{(\Delta_{\psi}p'_{j})_{\rm opt}}{(\Delta_{\psi}x'^{i})_{\rm opt}} \Delta_\psi x'^{i} = \frac{\tilde{\sigma}_{gj}}{\sigma_g^{i}}\Delta_\psi x'^{i} \, . 
\end{eqnarray}

\subsubsection{Generalised uncertainty principle} \label{GUP}

We now derive the GUP and argue that it is applicable in practically all situations of physical interest.
Recall the formula for smeared position uncertainty,
\begin{eqnarray} \label{Fund_Rel-GUP}
\Delta_\Psi X^{i} &=& \sqrt{ (\Delta_\psi x'^{i})^2 + (\sigma_g^{i})^2}
\nonumber\\
&\geq& \frac{\hbar}{2 \Delta_\psi p'_{i}} \sqrt{1 + \left(\frac{2\sigma_g^{i} \Delta_\psi p'_{i}}{\hbar}\right)^2} \, ,
\end{eqnarray}
where the inequality follows from HUP. 
Note that, here, there is no summation over index $i$. 
Instead, the inequality holds for the $i$th component of position vector.
The squared term inside the root is small if
\begin{eqnarray}
\Delta_\psi p'_{i} \ll \frac{\hbar}{2 \sigma_g^{i}} \, .
\end{eqnarray}
That is, practically always, as the right-hand side is the momentum uncertainty corresponding to an object localised to the Planck length.
In this case, expanding the square root to first order yields:
\begin{eqnarray} \label{GUP-final*}
\Delta_\Psi X^{i} \gtrsim \frac{\hbar}{2 \Delta_\psi p'_{i}} + \frac{(\sigma_g^{i})^2}{\hbar} \Delta_\psi p'_{i} \, .
\end{eqnarray}

From here on, we neglect dimensional indices, since the same relation holds in all orthogonal directions.
In three-dimension space, where $\sigma_g = \sqrt{2}l_{\rm Pl}$, we then have:
\begin{eqnarray} \label{GUP-final}
\Delta_\Psi X \gtrsim \frac{\hbar}{2 \Delta_\psi p'} + \frac{2G}{c^3} \Delta_\psi p' \, .
\end{eqnarray}
Finally, we note that $\Delta_\Psi X \simeq \Delta_\psi x'$ when $\Delta_\psi x' \gg \sqrt{2}l_{\rm Pl}$. 
Equation (\ref{GUP-final}) then takes the same form as Eq. (\ref{GUP-1*}), with $\alpha =2$, but with the heursistic uncertainties $\Delta x$ and $\Delta p$ replaced by the well defined standard deviations $\Delta_\psi x'$ and $\Delta_\psi p'$, respectively.

\subsubsection{Extended uncertainty principle} \label{EUP}

Similarly, we obtain the EUP by considering the smeared-space momentum uncertainty:
\begin{eqnarray} \label{Fund_Rel-EUP}
\Delta_\Psi P_{i} &=& \sqrt{(\Delta_\psi p'_{i})^2 + \tilde{\sigma}_{gi}^2} 
\nonumber\\
&\geq& \frac{\hbar}{2 \Delta_\psi x'^{i}} \sqrt{1 + \left( \frac{2 \Delta_\psi x'^{i} \, \tilde{\sigma}_{gi}}{\hbar} \right)^2} \, .
\end{eqnarray}
Again, note that here there is no summation over index $i$, and the inequality holds for the $i$th component of the momentum vector.
The squared term is small if
\begin{equation}
\Delta_\psi x'^{i} \ll \frac{\hbar}{2 \tilde{\sigma}_{gi}} \, ,
\end{equation}
which again holds in practically all situations of physical interest, as the limit on the right-hand side is of the order of the radius of the universe.
By expanding the square root to first order, we obtain:
\begin{eqnarray} \label{EUP-final}
\Delta_\Psi P_{i} 
\gtrsim \frac{\hbar}{2 \Delta_\psi x'^{i}} + \frac{(\tilde{\sigma}_{gi})^2}{\hbar} \Delta_\psi x'^{i} \, .
\end{eqnarray}

Again neglecting dimensional indices, and restricting ourselves to three spatial dimensions, gives:
\begin{eqnarray} \label{EUP-final}
\Delta_\Psi P 
\gtrsim \frac{\hbar}{2 \Delta_\psi x'} + \frac{\hbar \Lambda}{12} \Delta_\psi x' \, ,
\end{eqnarray}
For $\Delta_\psi p' \gg (1/2)m_{\rm dS}c$, we have $\Delta_\Psi P \simeq \Delta_\psi p'$, and Eq. (\ref{EUP-final}) takes the same form as Eq. (\ref{EUP-1}), with $\eta = 1/12$.

\subsubsection{Extended generalised uncertainty principle} \label{EGUP}

Note that, for both $\Delta_\Psi X^{i} \simeq \Delta_\psi x'^{i}$ and $\Delta_\Psi P_{j} \simeq \Delta_\psi p'_{j}$, taking the square root of Eq. (\ref{X_g_P_X_g_uncertainty-2}) and Taylor expanding the right-hand side yields the EGUP, also proposed in \cite{Bolen:2004sq,Park:2007az,Bambi:2007ty}, which reduces to both the GUP and EUP in appropriate limits. 
However, from our previous considerations, it is clear that both the GUP and EUP hold independently, in practically all situations of physical interest, irrespective of the EGUP. 

In other words, in the smeared-space formalism, the EGUP is not the fundamental uncertainty relation, from which the GUP and EUP are derived. 
Instead, the fundamental relations (\ref{Fund_Rel-GUP}) and (\ref{Fund_Rel-EUP}) give rise to the GUP and EUP, respectively, and may also be combined to give the EGUP. 
By contrast, in order to obtain both the GUP and the EUP from modified commutation relations we must modify the position-momentum commutator to first obtain the EGUP (see Eq. (\ref{modified_commutator-1})), before deriving the GUP and EUP as separate limits. 

This is one of several important differences between the smeared-space and modified commutator approaches to generalised uncertainty relations. 
Others will be discussed in the Conclusions, Sec. \ref{CONCLUSIONS}.

\subsection{Smeared-space wave mechanics} \label{SEC_WM}

We have introduced the generalised position operator, $\hat{X}^{i}$, and defined its action on the position space representation of the smeared-space wave functions as:
\begin{equation}
\hat{X}^{i} \Psi(\vec{x},\vec{x}{\, '}) = x'^{i} \Psi(\vec{x},\vec{x}{\, '}) \, .
\label{EQ_WM_POS}
\end{equation}
Similarly, we have introduced the generalised momentum operator, $\hat{P}_{i}$, and defined its action on the momentum space representation: 
\begin{equation}
\hat{P}_{i} \tilde \Psi(\vec{p},\vec{p}{\, '}) = p_{i}' \tilde{\Psi}(\vec{p},\vec{p}{\, '}) \, .
\label{EQ_WM_MOM}
\end{equation}
We now determine how to calculate generalised position and momentum statistics, without changing the representation of the measured state, by analogy with standard wave mechanics.
In the analysis that follows, it is helpful to recall the modified de Broglie relation, $\vec{p}{\, '} = \hbar \vec{k} + \beta \Delta_{\vec{k}'\vec{k}}$, where $\Delta_{\vec{k}'\vec{k}} = \vec{k}' - \vec{k}$, see Eq.~(\ref{mod_dB}).

We begin with the position space representation, where Eq. (\ref{mod_dB}) directly suggests the following form for $\hat{P}_{i}$:
\begin{eqnarray}
\hat{P}_{i} &=&  \hbar \hat{k}_{i} + \beta \hat \Delta_{k_{i}'k_{i}}
\nonumber\\
&=& - i \hbar \frac{\partial}{\partial x^{i}} - i \beta \frac{\partial}{\partial (x'^{i} - x^{i})} \, .
\end{eqnarray}
Indeed, one verifies that this gives the correct eigenvalue ($p'_{i}$) when applied to the position space representation of smeared-space momentum eigenstate, $\bra{\vec{x}}\braket{\vec{x}{\, '}|\vec{p} \, \vec{p}{\, '}}$ (\ref{smeared-p-eigenstate}).

We now consider the momentum space representation of $\hat{X}^{i}$. 
Just as the smeared-space momentum eigenvalue, $p'_{i}$, may be written as $p'_{i} = p_{i} + (p'_{i} - p_{i})$, where $p_{i}$ and $(p'_{i}-p_{i})$ act as independent variables, 
we may also decompose the smeared-space position eigenvalue, $x'^{i}$, as $x'^{i} = x^{i} + \Delta_{x'^{i}x^{i}}$, where $\Delta_{x'^{i}x^{i}} = x'^{i} - x^{i}$.
Accordingly, the generalised position operator in the momentum space representation is given by:
\begin{eqnarray}
\hat{X}^{i} &=& \hat{x}^{i} + \hat \Delta_{x'^{i}x^{i}}
\nonumber\\
&=& i \hbar \frac{\partial}{\partial p_{i}} + i \beta \frac{\partial}{\partial (p'_{i} - p_{i})} \, .
\end{eqnarray}
One verifies that this gives correct eigenvalue ($x'^{i}$) when applied to the momentum space representation of the smeared-space position eigenstate, $\braket{\vec{p} \, \vec{p}{\, '}|\vec{x}}\ket{\vec{x}{\, '}}$. 

Our previous considerations imply the modified free-particle Hamiltonian, $\hat{H} := \hat{P}^2/(2m)$, which acts on the smeared-state $\ket{\Psi}$, and we may conjecture that the position-basis expansion of the canonical potential operator,
\begin{eqnarray}
\hat{V} := \int V(\vec{x}) \ket{\vec{x}}\bra{\vec{x}} {\rm d}^d\vec{x} \, , 
\end{eqnarray}
should be mapped according to:
\begin{eqnarray}
\hat{V} \mapsto \hat{\mathcal{V}} := \hat{\openone} \otimes \hat{V}' \, , 
\end{eqnarray}
by analogy with $\hat{x}^{i} \mapsto \hat{X}^{i} := \hat{\openone} \otimes \hat{x}'^{i}$.
This suggests the modified Schr{\" o}dinger equation:
\begin{eqnarray} \label{mod_Schrod}
\hat{H}\ket{\Psi} = i(\hbar + \beta) \frac{d}{dt}\ket{\Psi} \, ,
\end{eqnarray}
where 
\begin{eqnarray} \label{gen_Hamiltonian}
\hat{H} := \frac{\hat{P}^2}{2m} + \hat{\mathcal{V}} \, ,
\end{eqnarray}
and $\ket{\Psi}$ is given by Eq. (\ref{EQ_PSIG}). 

Here, the substitution $\hbar \mapsto \hbar + \beta$ on the right-hand side of Eq. (\ref{mod_Schrod}) is suggested by the form of the smeared-space position-momentum commutator, Eq. (\ref{[X,P]}), together with
\begin{eqnarray} \label{[H,X]}
[\hat{H},\hat{X}^{i}] = \frac{(\hbar + \beta)}{i}\frac{\hat{P^{i}}}{m} 
\end{eqnarray}
and
\begin{eqnarray} \label{[H,X]}
[\hat{H},\hat{P}_i] = [\hat{\mathcal{V}},\hat{P}_{i}]  = -\frac{(\hbar + \beta)}{i}\frac{\partial \hat{\mathcal{V}}}{\partial x'^{i}} \, . 
\end{eqnarray}
This, in turn, suggests the modified energy-frequency de Broglie relation:
\begin{eqnarray} \label{E_omega_mod_dB}
E = (\hbar + \beta)\omega \, ,  
\end{eqnarray}
and, hence, the modified quantum dispersion relation for a free particle on the smeared-space background:
\begin{eqnarray} \label{mod_disp_rel}
\omega = \frac{(\hbar \vec k + \beta(\vec k' - \vec k))^2}{2m(\hbar+\beta)} \, . 
\end{eqnarray}

The time-dependent smeared-state then takes the form: 
\begin{eqnarray} \label{time-evol}
\ket{\Psi_{t}} = \hat{U}(t)\ket{\Psi_{0}} \, , 
\end{eqnarray}
where $\ket{\Psi_{0}} \equiv \ket{\Psi}$ and, when $\hat{H}$ is independent of $t$, the unitary time-evolution operator is given by
\begin{eqnarray} \label{U(t)}
\hat{U}(t) = e^{-\frac{i}{(\hbar + \beta)}\hat{H}t}  \, .
\end{eqnarray}
The associated Heisenberg equation is:
\begin{eqnarray} \label{mod_Heisenberg}
\frac{d}{dt} \hat{O}(t) = \frac{i}{(\hbar + \beta)}[\hat{H},\hat{O}] + \frac{\partial \hat O}{\partial t} \, ,
\end{eqnarray}
where both $\hat{O}$ and $\hat{H}$ act on the tensor product space. 

We note that, since, in canonical QM, time is a parameter and not an operator, there are no superpositions in $t$, no $t$-eigenvalues, and no kets $\ket{t}$. 
Hence, we cannot `smear' time, by analogy with the smearing of space: that is, by introducing an extra degree of freedom $t'$. 
Nonetheless, since the time evolution of $\ket{\Psi}$ is generated by the Hamiltonian $\hat{H}$ (\ref{gen_Hamiltonian}), the time evolution of the canonical state $\ket{\psi}$ should be generated by its appropriately smeared counterpart in the smeared-operator picture (see Sec.~\ref{smeared_ops}). 
Therefore, it is clear that the time evolution of the canonical state $\ket{\psi}$ is, in some sense, `smeared', even though time itself is not. 

\subsection{Multi-particle states} \label{MULTI}

We conclude this section with a brief description of multi-particle states. 
The construction of $n$-particle wave functions in the smeared-space formalism is potentially non-trivial, since, for $n > 1$, the canonical wave function ceases to be a function on real (physical) space, and is instead defined on an $(n \times d)$-dimensional configuration space in both the position and momentum space representations. 
Hence, we must consider carefully how to `smear' configuration space for $n > 1$.

We begin by noting that, in canonical QM, the configuration space of $n$ particles in one dimension is equivalent to the configuration space of a one-particle state in $n$ dimensions. 
In both cases, one simply makes the following transition from the one-dimensional one-particle state: $x \rightarrow \vec{\xi}$, ${\rm d}x \rightarrow {\rm d}^n\vec{\xi}$, $\psi(x) \rightarrow \psi(\vec{\xi})$, where $\vec{\xi}$ is an $n$-dimensional vector. 
In the former case, this represents the $n$ independent coordinates of the position of a single particle, whereas, in the latter, it represents the one-dimensional positions of $n$ separate particles. 
Likewise, $n$-particle states in $d$ dimensions may be constructed by making the following transition from the $d$-dimensional one-particle state: $\vec{x} \rightarrow \vec{\xi}$, ${\rm d}^d\vec{x} \rightarrow {\rm d}^{d \times n}\vec{\xi}$ and $\psi(\vec{x}) \rightarrow \psi(\vec{\xi})$, where $\vec{\xi} = (\vec{x}_1,\vec{x}_2, \dots ,\vec{x}_n)$.

In full analogy to the canonical theory, we therefore construct $n$-particle wave functions in the smeared-space formalism as:
\begin{eqnarray}
\Psi(\vec{\xi},\vec{\xi}{\, '}) := g(\vec{\xi}{\, '}-\vec{\xi}) \psi(\vec{\xi}) \, , 
\end{eqnarray}
in the position space representation, where $\vec{\xi}{\, '} = (\vec{x}_1{\, '},\vec{x}_2{\, '}, \dots ,\vec{x}_n{\, '})$ and $g(\vec{\xi}{\, '}-\vec{\xi})$ is a normalised function of $|\vec{\xi}{\, '}-\vec{\xi}|$. 
The state $\ket{\Psi}$ is then given by:
\begin{eqnarray} \label{multi-particle}
\ket{\Psi} &:=& \int \int g(\vec{\xi}{\, '}-\vec{\xi}) \psi(\vec{\xi}) \ket{\vec{\xi}} \otimes \ket{\vec{\xi}{\, '}} {\rm d}^{n \times d}\vec{\xi}{\rm d}^{n \times d}\vec{\xi}{\, '}
\nonumber\\
&=& \int \dots \int g(\vec{x}_1{\, '}-\vec{x}_1, \dots ,\vec{x}_n{\, '}-\vec{x}_n)\psi(\vec{x}_1, \dots ,\vec{x}_n)
\nonumber\\
&{}&  \bigotimes_{l=1}^{n} \ket{\vec{x}_l} \ket{\vec{x}_l{\, '}} \prod_{l=1}^{n}{\rm d}^{d}\vec{x}_{l}{\rm d}^{d}\vec{x}_{l}{\, '} \, . 
\end{eqnarray}
On the first line of Eq. (\ref{multi-particle}), each integral represents integration over an $(n \times d)$-dimensional subspace of the $(2n \times d)$-dimensional configuration space of the smeared $n$-particle state. 
On the second line, we envisage $2n$ integrals, each of which integrates over one of the $d$-dimensional subspaces represented by ${\rm d}^{d}\vec{x}_{l}$ or ${\rm d}^{d}\vec{x}_{l}{\, '}$, for fixed $l$.

The momentum space representation of multi-particle states then follows by full analogy with the momentum space representation of smeared one-particle states (see Sec.~\ref{MOMENTUM}) and we may define an appropriate multi-particle smearing operator, $\hat{S}$, that acts on canonical multi-particle states $\ket{\psi}$ in the smeared-operator picture. 
The position and momentum observables, $\hat{X}^{i}$ and $\hat{P}_{i}$, are also extended in a natural way, by analogy with the multi-particle extensions of $\hat{x}^{i}$ and $\hat{p}_{i}$, combined with our previous results.

In the context of multipartite systems, an interesting question emerges, namely, whether smearing can induce entanglement between spatially separated particles.
The answer depends on the form of the smearing function.
Consider, for simplicity, two particles on the real line, i.e. in a toy one-dimensional universe.
If the particles are in a product state in the fixed-background, their wave function factorises:
\begin{equation}
\psi_{12} (x_1, x_2) = \psi_1(x_1) \psi_2(x_2) \, .
\end{equation}
Applying the smearing procedure described above then produces the smeared wave function
\begin{equation}
\Psi_{12} (x_1, x_2,x_1',x_2') = g(\vec \xi{\, '} - \vec \xi) \psi_1(x_1) \psi_2(x_2) \, , 
\end{equation}
where $\vec{\xi} = (x_1,x_2)$.
If the smearing function factorises, as is the case for Gaussian smearing, we obtain
\begin{eqnarray}
\Psi_{12} (x_1, x_2,x_1',x_2') & = & g(x_1' - x_1) g(x_2' - x_2)  \psi_1(x_1) \psi_2(x_2) \nonumber \\
& = & \Psi(x_1,x_1') \Psi(x_2,x_2').
\end{eqnarray}
Therefore, unentangled particles in fixed background are also unentangled in smeared space.
However, for non-Gaussian smearing functions that do not factor as above, smearing can induce entanglement.

\section{Applications} \label{SEC_APPLICATIONS}

In this section, we apply the smeared-space formalism developed in Sec.~\ref{SEC_FORMALISM} to outstanding problems in cosmology and astrophysics. 
Thus, we restrict our attention, from here on, to three spatial dimensions, and to the observed values of $\rho_{\Lambda}$ and $\rho_{\rm Pl}$. 
In Sec.~\ref{Cosmology}, we show that an object in smeared-space, described by a wave function $\Psi$ that optimises the lower bound on the product of uncertainties, $\Delta_{\Psi}X\Delta_{\Psi}P$, has an energy density of order $\sim \rho_{\Lambda}$. 
We then discuss possible observational consequences of the smeared-space uncertainty relations (\ref{GUR_X})-(\ref{GUR_P}) and their implications for the nature of dark energy. 
In Sec.~\ref{BHUP}, we show how the GUP derived from Eq. (\ref{GUR_X}) may be tentatively extended into the black hole regime, yielding a concrete realisation of the black hole uncertainty principle (BHUP) correspondence conjectured in \cite{Carr:2014mya}. 
Finally, in Sec.~\ref{FiniteU}, we consider the nature of generalised uncertainty relations in a finite universe. 
We argue that a more thorough treatment of this problem, including finite-horizon effects, implies both maximum and minimum bounds on the generalised position and momentum uncertainties, resulting in stronger constraints than those obtained in Sec.~\ref{SEC_FORMALISM}.
This may be regarded as a limitation of the present formalism and we briefly outline the steps required to extend it to the more general case.

In contrast to the precise mathematical statements that define the smeared-space formalism, the discussion presented here is, necessarily, more speculative in nature. 
Since it concerns important open problems in fundamental physics, such as the nature of dark energy and the quantum description of black holes, this is largely unavoidable. 
Nonetheless, we include even speculative arguments, since the implications of the smeared-space model for fundamental physics, including its connections to existing theories, and to empirical data, have yet to be explored in detail. 
For the sake of notational simplicity, we neglect dimensional indices throughout this section, unless they are explicitly required.

\subsection{Cosmology} \label{Cosmology}

We now focus our attention on the optimum position and momentum uncertainties, $(\Delta_\psi x')_{\rm opt}$ (\ref{EQ_CAN_DX_OPT}) and $(\Delta_\psi p')_{\rm opt}$ (\ref{EQ_CAN_DP_OPT}), 
which minimise the product of the generalised uncertainties, $\Delta_\Psi X \, \Delta_\Psi P$ (\ref{X_g_P_X_g_uncertainty-2}). 
Substituting for $\sigma_g$ and $\tilde{\sigma}_g$ from Eq. (\ref{sigma_g}), these may be rewritten as
\begin{eqnarray} \label{}
(\Delta_\psi x')_{\rm opt} = l_{\Lambda} \, , \quad (\Delta_\psi p')_{\rm opt} = \frac{1}{2}m_{\Lambda}c \, , 
\end{eqnarray}
where
\begin{eqnarray} \label{}
l_{\Lambda} := 2^{1/4}\sqrt{l_{\rm Pl}l_{\rm dS}} \simeq 0.1 \ {\rm mm} \, , 
\end{eqnarray}
and
\begin{eqnarray} \label{}
m_{\Lambda} := 2^{-1/4}\sqrt{m_{\rm Pl}m_{\rm dS}} \simeq 10^{-3} \ {\rm eV} \, . 
\end{eqnarray}
Though extremely small compared to typical macroscopic length- and mass-scales, we will now show how these scales are relevant to cosmology.

In \cite{Mak:2001gg,Boehmer:2005sm} it was shown that, in the presence of dark energy, a spherically symmetric compact object must have a mean density greater than or approximately equal to the dark energy density in order to remain stable.  
Although the proof of this statement is non-trivial, requiring the use of the generalised Buchdahl inequalities in general relativity \cite{Buchdahl:1959zz}, its physical reason is intuitively clear, since bodies with $\rho \lesssim \rho_{\Lambda}$ have insufficient self-gravity to overcome the effects of dark energy repulsion. 

Thus, defining the mass density associated with the Compton radius of a particle as
\begin{eqnarray} \label{}
\rho_{\rm part} := \frac{3}{4\pi}\frac{m}{\lambda_{\rm C}^3(m)} \, , 
\end{eqnarray}
and requiring $\rho_{\rm part} \geq \rho_{\Lambda} = \Lambda c^2/(8\pi G)$ (\ref{DE_density}), implies $m \geq m_{\Lambda}$ ($\lambda_{\rm C}(m) \leq l_{\Lambda}$).
The scale $m_{\Lambda}$ may therefore be interpreted as the minimum possible rest mass of a stable, compact, charge-neutral, self-gravitating and quantum mechanical object, in the presence of dark energy \cite{Burikham:2015nma}.
It is interesting to note that this is comparable to the current bound on the mass of the electron neutrino, the lightest particle of the standard model, obtained from Planck satellite data \cite{PlanckCollaboration}.

Furthermore, we note that the wave packet of a photon, or of an ultra-relativistic massive particle, will have an energy density comparable to the dark energy density when it is localised to a sphere of radius $\Delta_\psi x' \simeq (\Delta_\psi x')_{\rm opt}$ and has a momentum uncertainty of order $\Delta_\psi p' \simeq (\Delta_\psi p')_{\rm opt}$, i.e.,
\begin{eqnarray} \label{}
\mathcal{E}_\psi \simeq \frac{3}{4\pi}\frac{(\Delta_\psi p')_{\rm opt}\ c}{(\Delta_\psi x')^3_{\rm opt}} \simeq \rho_{\Lambda}c^2 \, . 
\end{eqnarray}
By contrast, the wave functions of non-relativistic particles of mass $m$ have energy densities comparable to $\rho_{\Lambda}$ when $\Delta_\psi x' \simeq (\Delta_\psi x')_{\rm opt}$ and $\Delta_\psi p' \simeq \sqrt{(\Delta_\psi p')_{\rm opt}mc}$. This is most naturally realised for $m \simeq m_{\Lambda}$.

This observation suggests a granular model of dark energy in which, whatever its underlying nature or dynamics, the dark energy field remains trapped in a Hagedorn-type phase \cite{Burikham:2015nma}.
In this scenario, there exists a space-filling `sea' of fermionic dark energy particles, each of mass $m_{\Lambda} \simeq 10^{-3}$ eV, with an average inter-particle distance of $\lambda_{\rm C}(m_{\Lambda}) = l_{\Lambda} \simeq 0.1$ mm.
Hence, any attempt to further reduce the distance between a pair of neighbouring particles, even if this results from random quantum fluctuations implied by the uncertainty principle, leads to the pair-production of new particles, rather than an increase in average energy density. 
Since space is already `full', carrying the critical (Hagedorn) density of dark energy particles, new particles cannot be created without a concomitant expansion of space itself, leading to the accelerated expansion of the universe \cite{Burikham:2015nma}.

This model has a number of attractive features. 
First, it requires a pair-production rate of the order of one pair per de Sitter volume, $\sim l_{\rm dS}^3$, per Planck time, $t_{\rm Pl} = l_{\rm Pl}/c$, in order to give rise to the present rate of expansion, which is inferred from type 1a supernovae data
\cite{Reiss1998,Perlmutter1999}, observations of large-scale structure \cite{Betoule:2014frx}, and the cosmic microwave background (CMB) radiation \cite{PlanckCollaboration}.
In other words, if a single pair of dark energy particles, each of radius $\sim 0.1$ mm, is created somewhere in the observable universe every $\sim10^{-43}$ seconds, galaxies will recede from one another at the observed Hubble rate \cite{Burikham:2015nma}.

Second, $m_{\Lambda}$ is the unique mass-scale for which the Compton wavelength of a particle is equal to its gravitational turn-around radius, i.e., the radius at which dark energy repulsion overcomes canonical (Newtonian) gravitational attraction \cite{Bhattacharya:2016vur}.
This gives a neat interpretation of the stability condition $\rho_{\rm part} \geq \rho_{\Lambda}$ and suggests that $m_{\Lambda}$ is the unique scale for which the (positive) rest mass of a body is counter-balanced by its (negative) gravitational energy \cite{Burikham:2017bkn,Lake:2017ync,Lake:2017uzd}.
In this way, particles with rest mass $m_{\Lambda}$ can be pair-produced ad infinitum, leading to eternal universal expansion and the existence of an asymptotic de Sitter phase, $a(\tau) \propto e^{\sqrt{\Lambda/3}c\tau}$, where $\tau$ is the cosmic time and $a(\tau)$ is the scale factor of the universe. 
A model of this form was first proposed in \cite{Burikham:2015nma}, though a more detailed model of universal expansion from eternal fermion production was recently proposed in \cite{Hashiba:2018hth}.  

We note that, in this model, we expect the dark energy field to exhibit granularity over a length-scale of order $\sim 0.1$ mm, while remaining approximately constant over much larger scales. 
Specifically, taking $k_{\rm max} \simeq 2\pi/l_{\Lambda}$ as the not-so-UV cut-off for vacuum field modes yields a vacuum energy density of order
\begin{eqnarray} \label{rho_vac}
\rho_{\rm vac} &\simeq& \frac{\hbar}{c} \int_{2\pi/l_{\rm dS}}^{2\pi/l_{\Lambda}} \sqrt{k^2 + \left(\frac{2\pi}{l_{\Lambda}}\right)^2} d^3k 
\nonumber\\ 
&\simeq& \frac{\Lambda c^2}{G} \simeq 10^{-30} \ {\rm g \ . \ cm^{-3}} \, ,
\end{eqnarray}
as required. Here, modes with $k > 2\pi/l_{\Lambda}$ immediately stimulate the pair-production of dark energy particles \cite{Lake:2017ync,Lake:2017uzd}, triggering universal expansion in place of increased energy density, as described above.  

With this in mind, it is intriguing that tentative observational evidence for the periodic variation of the gravitational field strength on a length-scale of order $l_{\Lambda} \simeq 0.1$ mm has recently been proposed, though, at present, the confidence level is no more than $2\sigma$ \cite{Perivolaropoulos:2016ucs,Antoniou:2017mhs}.
Although various models of modified (non-Einstein) gravity predict such spatial periodicity in the low-energy `Newtonian' regime (see \cite{Perivolaropoulos:2016ucs} and references therein), it is certainly consistent with the granular dark energy models proposed in \cite{Burikham:2015nma,Burikham:2017bkn,Lake:2017ync,Lake:2017uzd,Hashiba:2018hth}. 
It is striking that the same length-scale appears naturally by optimising the uncertainty relations derived from the smeared-space formalism, Eqs. (\ref{GUR_X})-(\ref{GUR_P}).

From a cosmological perspective, another intriguing aspect of the smeared-space model is that, since both $\sigma_g$ and $\tilde{\sigma}_g$, which are identified with the Planck length and de Sitter momentum, respectively, are required to be finite and strictly positive, it is impossible to construct a consistent theory with minimum length $\sim l_{\rm Pl}$ without introducing a minimum momentum $\sim m_{\rm dS}c$.
This, in turn, implies the existence of a maximum horizon distance of order $\sim l_{\rm dS}$, and, hence, a  minimum energy density $\rho_{\Lambda} = \Lambda c^2/(8\pi G)$, with $\Lambda > 0$.
In other words, the existence of a positive dark energy density is logically necessary, in the smeared-space model, 
since the quantisation of physical space implies a concomitant quantisation of momentum space. 
Thus, though optional in classical general relativity, the arguments presented here suggest that a universe with no dark energy ($\Lambda = 0$) would be inconsistent at the quantum level. 
(Specifically, we recall that $\Lambda > 0$ is required in order to maintain the basis independence of the smeared state $\ket{\Psi}$, (\ref{EQ_PSIG}) and (\ref{EQ_PSIG*}).)
The same argument rules out the physical existence of anti-de Sitter space ($\Lambda < 0$), since $l_{\rm dS} = \sqrt{3/\Lambda}$ and $\tilde{\sigma}_g \simeq \hbar\sqrt{\Lambda/3}$ are, of course, required to be real.

Furthermore, several theoretical and observational studies in the recent literature suggest the relevance of the scales $l_{\Lambda} \simeq 0.1 \, {\rm mm}$ and $m_{\Lambda} \simeq 10^{-3}$ eV to cosmology and high-energy physics, in a variety of contexts. 
In \cite{Harko:2015aya}, galactic radii data and observational constraints from the bullet cluster collision were used to determine the mass, $m_{\chi}$, of a candidate Bose-Einstein condensate dark matter particle, yielding an estimate of order  $m_{\chi} \simeq 10^{-2}-10^{-4}$ eV. 
In \cite{Ong:2018nzk}, it was shown that the EGUP, which may be obtained by Taylor expanding the square root of Eq. (\ref{X_g_P_X_g_uncertainty-2}) to first order, preserves the standard expression for the Chandrasekhar limit when applied to neutron stars, in contradistinction to the GUP.
Thus, theories with minimum length- and momentum-scales were shown to be consistent with astrophysical observations of massive compact objects, whereas theories with only a minimum length-scale may contradict existing data. 
In addition, according to the action uncertainty principle \cite{Garay:1994en}, $\Delta l \simeq \sqrt{l_{\rm Pl}l}$ represents the minimum uncertainty inherent in a measurement of the length-scale $l$ due to quantum gravity effects. 
In this interpretation, $l_{\Lambda}$ represents the minimum possible uncertainty in a measurement of the horizon distance $\sim l_{\rm dS}$.

A recent $F$-theory approach to the cosmological constant problem \cite{Heckman:2018mxl} also suggests a split mass spectrum for superpartners of order $\Delta M \sim \sqrt{M_{\rm UV}M_{\rm IR}}$, where $M_{\rm UV}$ and $M_{\rm IR}$ denote the ultraviolet and infrared cut-offs of the model, respectively. 
With reference to the smeared-space formalism, this result is particularly interesting, since the standard model contains two massless spin-1 bosons: the photon and the gluon. 
A massless spin-2 boson, the `graviton', has also been proposed, at least as an effective description of quantum gravity in the linearised gravity regime \cite{Pauli-Fierz}. 
Thus, in such a model, fermionic dark energy particles of mass $m_{\Lambda} \simeq \sqrt{m_{\rm Pl}m_{\rm dS}}$ may be the superpartners of the force-mediating bosons of the gravitational field, the electroweak force, or, in principle, even the strong nuclear force. 
 
While the first may seem the most natural, homogeneous and isotropic configurations of massive fields with spin $\geq 2$ are believed to be unstable, in both general relativity and modified gravity theories \cite{DeFelice:2012mx}, leading to instabilities in the cosmological solutions of the field equations. 
The second is plausible but surprising, in that it implies an intimate connection, not only between the macroscopic and microscopic worlds, but between the very essence of `dark' and `light' physics \cite{Lake:2017ync}.
Such a connection was postulated in \cite{Nottale,Boehmer:2006fd,Beck:2008rd,Burikham:2015sro,Lake:2017ync,Lake:2017uzd} as a physical explanation for the numerical coincidence $\Lambda \simeq m_{e}^6G^2/(\alpha_{e}^6\hbar^4) \simeq 10^{-56} \, {\rm cm^{-2}}$, where $m_{e}$ is the electron mass and $\alpha_{e} = e^2/(\hbar c)$ is the electromagnetic fine-structure constant. 
However, to date, there is no empirical evidence to support this.
Finally, the third seems the least plausible, since gluons (and hence gluinos) carry colour charge, and are expected to interact strongly with nuclear matter \cite{Donoghue:1992dd}, so that such effects should already have been observed.
 
Hence, at present, it is not clear how the smeared-space model is related to other candidate theories of quantum gravity. 
However, it is noteworthy that it shares a number of common features with independent studies. 
In particular, at least two approaches considered in the recent literature also involve a doubling of the classical gravitational degrees of freedom. 
The first, based on a self-dual action for a non-commutative geometry in loop quantum gravity, involves a doubling of the tetrad degrees of freedom in canonical general relativity \cite{deCesare:2018cjr}.
The second, based on the holographic quantisation of higher-spin gravity on a de Sitter causal patch, explicitly utilises the tensor product construction $\mathcal{H} \otimes \mathcal{H}^{*}$, together with a transformation to light-cone coordinates $(x^{+},x^{-})$ \cite{Neiman:2018ufb}. 
In the smeared-space model, the physical meaning of this transformation is clear.
If the metric on the $(x,x')$-plane is Minkowski (see Sec. \ref{CONCLUSIONS}), $x^{-} := (1/2)(x'-x)$ represents the space-like direction in the smeared geometry, which is parallel to the most probable universe (i.e., the diagonal line in the one-dimensional example presented in Fig.~\ref{FIG_PHASESPACE}) and $x^{+} := (1/2)(x'+x)$ represents the orthogonal time-like direction.  

Finally, we note that, from a cosmological perspective, our procedure for the smearing of momentum space, presented in Sec.~\ref{MOMENTUM}, cannot be regarded as fundamental. 
Since our rationale for the introduction of a minimum momentum-scale was the existence of a maximum length-scale (i.e., the de Sitter radius), we note that, prior to the present epoch, the radius of the universe was much smaller than the de Sitter horizon.
Heuristically, this suggests the replacement:
\begin{eqnarray} \label{agegraphic-1}
\tilde{\sigma}_g := \frac{1}{2}m_{\rm dS}c \rightarrow \tilde{\sigma}_g(\tau) := \frac{1}{2}m_{\rm \mathcal{H}}(\tau)c \, , 
\end{eqnarray}  
where
\begin{eqnarray} \label{agegraphic-2}
m_{\rm \mathcal{H}}(\tau) := \frac{\hbar}{l_{\rm \mathcal{H}}(\tau)c} \, , \quad l_{\rm \mathcal{H}}(\tau) := \frac{c}{\mathcal{H}(\tau)} \, . 
\end{eqnarray} 
Here, $\mathcal{H}(\tau) = \dot{a}/a$ is the Hubble parameter, and $l_{\mathcal{H}}(\tau)$ is of the order of the cosmological horizon at time $\tau$. 
Therefore \cite{Lake:2017ync,Lake:2017uzd}:
\begin{eqnarray} \label{agegraphic-3}
&&(\Delta_{\psi}p')_{\rm opt} \simeq \sqrt{m_{\rm Pl}m_{\rm dS}}c 
\nonumber\\
&\rightarrow& (\Delta_{\psi}p')_{\rm opt}(\tau) \simeq \sqrt{m_{\rm Pl}m_{\mathcal{H}}(\tau)}c \, . 
\end{eqnarray} 

This suggests an agegraphic model of dark energy, similar to those proposed in \cite{Cai:2007us,Wei:2007ty}, which reduces approximately to the $\Lambda$CDM concordance model only at the present epoch, where $\mathcal{H}(0)/c \simeq \sqrt{\Lambda/3}$. 
However, such a macro-model cannot easily be identified with the pair-production of fermionic dark energy particles at the micro-level, since the existence of a time-dependent rest mass implies violation of Lorentz invariance, and, hence, of energy and momentum conservation.
Nonetheless, identifying $\sqrt{m_{\rm Pl}m_{\mathcal{H}}(\tau)}$, instead, with the (energy-dependent) renormalised mass of the dark energy fermions, the two pictures may be reconciled.
Equation (\ref{agegraphic-3}) then suggests a novel form of unification at the big bang, since, for $l_{\mathcal{H}}(\tau) \rightarrow l_{\rm Pl}$, we have $m_{\rm \mathcal{H}}(\tau) \rightarrow m_{\rm Pl}$. 
In this scenario, the renormalised mass of the lightest fundamental particle converges to the upper limit for all particles, suggesting a unification of all particle masses and fundamental forces \cite{Lake:2017ync,Lake:2017uzd}.

In the smeared-space model, the $\tau \rightarrow 0$ limit implied by agegraphic theories (i.e., $\rho_{\rm vac} \rightarrow \rho_{\rm Pl}$) is particularly interesting, since it implies $\beta \rightarrow \hbar$.
In this limit, the physical momentum $p'$ (\ref{mod_dB}) is a function of $k'$ only, and the uncertainty principle for spatial `points', Eq. (\ref{EQ_BETA_SCALE}), is equivalent to the HUP. 
Naively, this suggests the elimination of the distinction between matter living `in' a geometry, and the quantum state of the geometry itself, though one must be cautious when extrapolating formulae, such as Eq. (\ref{EQ_BETA_SCALE}), so far beyond their expected region of validity. 

Nonetheless, we note that, in principle, the smearing scale for momentum space may remain fixed ($\tilde{\sigma}_g \simeq m_{\rm dS}c$) throughout the cosmological history, while the minimum momentum in each classical background geometry varies as $(\Delta_{\psi} p)_{\rm min}(\tau) \simeq m_{\mathcal{H}}(\tau)c$.
In this scenario, the two values coincide only at the present epoch, as the universe undergoes the transition from a deccelerating phase to an asymptotically de Sitter expansion.

\subsection{The BHUP correspondence} \label{BHUP}

In this section, we consider a possible relation between the GUP, as formulated in the smeared-space model, and the black hole uncertainty principle (BHUP) correspondence, proposed in \cite{Carr:2014mya}.
We recall that the BHUP correspondence posits the existence of a unified expression for the radii of black holes and fundamental particles, and, for this reason, is also referred to as the Compton-Schwarzschild correspondence \cite{Lake:2015pma,Lake:2016did,Lake:2016enn,Lake:2018hyv}.

Though fundamentally a result of relativistic quantum theory (i.e., quantum field theory), the standard expression for the reduced Compton wavelength of a particle of mass $m$, 
\begin{eqnarray} \label{Compton}
\lambda_{\rm C}(m) = \frac{\hbar}{mc} \, , 
\end{eqnarray}
may also be obtained, heuristically, in non-relativistic quantum mechanics. 
Substituting the limit $(\Delta_\psi p)_{\rm max} \simeq mc$ into the HUP yields $(\Delta_\psi x)_{\rm min} \simeq \hbar/(2mc) \simeq \lambda_{\rm C}(m)$.
The physical intuition behind this result is that wave packets with momentum uncertainty $\Delta_\psi p \gtrsim mc$ have sufficient energy to pair-produce particles of mass $m$.
Thus, further increment in momentum results in the pair-production of new particles rather than increased localisation of the single-particle wave function. 
By contrast, in the gravitational regime of the mass-radius diagram \cite{Carr:2014mya}, the radius of a classical point-mass is the Schwarzschild radius,
\begin{eqnarray} \label{Schwarzschild}
r_{\rm S}(m) = \frac{2Gm}{c^2} \, . 
\end{eqnarray}

It is noteworthy that, substituting $\Delta_\psi p' \simeq mc$ into the GUP (\ref{GUP-final}) and identifying $\Delta_\Psi X \simeq \Delta_\psi x' \equiv R_{\rm C/S}(m)$, we obtain the unified expression
\begin{eqnarray} \label{R_C/S}
R_{\rm C/S}(m) \gtrsim \frac{\hbar}{2mc} + \frac{2Gm}{c^2} \, , 
\end{eqnarray}
in which the scale $R_{\rm C/S}(m)$ reduces approximately to the standard expressions for the Compton and Schwarzschild radii in the limits $m \ll m_{\rm Pl}$ and $m \gg m_{\rm Pl}$, respectively. 
However, we also note that, in the formalism presented here, Eq. (\ref{GUP-final}) 
is valid only within the momentum range $(1/2)m_{\rm dS}c \leq \Delta_\psi p' \leq (1/2\sqrt{2})m_{\rm Pl}c$, which corresponds to the fundamental particle region of the mass-radius diagram and explicitly excludes the black hole sector \cite{Carr:2014mya}. 
This is because the smeared-space formalism represents a generalisation of the canonical quantum formalism for fundamental particles, which is valid only within the sub-Planck mass domain.

Thus, although it is tempting to think that the identification $\Delta_\psi p' \simeq mc$ gives rise to a concrete realisation of the BHUP correspondence, this is not the case.
Nonetheless, we may construct a physical argument that allows us to tentatively extend the expression (\ref{R_C/S}) beyond the usual quantum regime, utilising the GUP (\ref{GUP-final}).

Consider, for the sake of simplicity, a black hole in the classical-background theory of canonical QM. 
Initially, the black hole is at rest in our chosen coordinate system, before emitting a particle via Hawking radiation.
Classically, $p_i^{\rm tot} = p_i^{\rm bh} + p_i^{\rm part} = 0$ along the line of particle emission, which we label as the coordinate direction $x^i$, due to momentum conservation. 
From Ehrenfest's theorem \cite{Griffiths} we have the same relation for the quantum mechanical expectation values of the momenta of well-localised objects, i.e. $\langle \hat p_i^{\rm bh} \rangle_\psi = - \langle \hat p_i^{\rm part} \rangle_\psi$,
where the expectation values are calculated for a suitable subsystem of the two-body state $\ket{\psi}$.
Furthermore, since the emission is spherically symmetric, $\braket{p_i^{\rm bh}}_\psi = \braket{p_i^{\rm part}}_\psi = 0$ holds in any direction $x^{i}$, 
yielding $\Delta_\psi p_i^{\rm bh} = \sqrt{\braket{(p_i^{\rm bh})^2}_\psi}$ and $\Delta_\psi p_i^{\rm part} = \sqrt{\braket{(p_i^{\rm part})^2}_\psi }$.

Next, we note that $\langle (\hat{p}_i^{\rm bh})^2 - (\hat{p}_i^{\rm part})^2 \rangle_\psi = \langle (\hat{p}_i^{\rm bh} + \hat{p}_i^{\rm part})(\hat{p}_i^{\rm bh} - \hat{p}_i^{\rm part}) \rangle_\psi$. 
This follows from the fact that $[\hat{p}_i^{\rm bh},\hat{p}_i^{\rm part}] = 0$, since $\hat{p}_i^{\rm bh}$ and $\hat{p}_i^{\rm part}$ represent local measurements on spatially isolated subsystems.
Assuming that the momentum of the center of mass is uncorrelated with the relative momenta, i.e. 
$\langle (\hat{p}_i^{\rm bh} + \hat{p}_i^{\rm part})(\hat{p}_i^{\rm bh} - \hat{p}_i^{\rm part}) \rangle_\psi = \langle \hat{p}_i^{\rm bh} + \hat{p}_i^{\rm part} \rangle_\psi \langle \hat{p}_i^{\rm bh} - \hat{p}_i^{\rm part} \rangle_\psi$,
we finally obtain $\braket{(p_i^{\rm bh})^2}_\psi = \braket{(p_i^{\rm part})^2}_\psi$, and, hence, $\Delta p_i^{\rm bh} = \Delta p_i^{\rm part}$.
The statistical spread of black hole recoil momenta is therefore equal to the statistical spread of the momenta of emitted particles, along any line of sight $x^i$, as expected intuitively. 
(From here on, we again neglect dimensional indices.)

However, it is well known that black holes of mass $M$ emit particles with typical masses $m \lesssim m_{\rm Pl}^2/M$, or energies $E \lesssim (m_{\rm Pl}^2/M)c^2$ in the case of massless particle emission \cite{Belgiorno:2019ofm}.
This follows from the requirement that the Compton (or de Broglie) wavelength of the emitted particle must be larger than or approximately equal to the Schwarzschild radius, $\lambda_{\rm C}(m), \, \lambda \gtrsim r_{\rm S}(M)$, in order for the particle to `escape' from the black hole.  
Thus, black hole recoil, and the corresponding conservation of momentum, suggest the following identifications in the gravitational region of the mass-radius diagram:
\begin{eqnarray} \label{alt_ident}
\Delta_{\psi}p^{\rm bh} = \Delta_{\psi}p^{\rm part} \simeq mc \lesssim \frac{m_{\rm Pl}^2}{M} c \, ,
\end{eqnarray}
where $m$ denotes that mass of the emitted quantum particle and $M$ is the black hole mass.

Switching to the smeared-space picture and identifying $\Delta_{\Psi}X \simeq \Delta_{\psi}x'^{\rm bh} \simeq \Delta_{\psi}x'^{\rm part}$ implies $m \simeq (1/4)(m_{\rm Pl}^2/M)$ and, hence, 
$\Delta_{\psi} p'^{\rm bh} \simeq \Delta_{\psi} p'^{\rm part} \simeq mc \simeq (1/4)(m_{\rm Pl}^2/M)c$.
Defining  $\Delta_{\psi} x'^{\rm bh} \equiv R_{C/S}(M)$, and substituting the above values into Eq. (\ref{GUP-final}), we obtain the following expression for the radius of a super-Planck mass `particle', i.e., a black hole:
\begin{eqnarray} \label{R_C/S-2}
R_{\rm C/S}(M) \gtrsim \frac{2GM}{c^2} + \frac{\hbar}{2Mc} \, .
\end{eqnarray}
This expression, which is valid for $M \gtrsim m_{\rm Pl}$, represents the generalised event horizon postulated in \cite{Carr:2014mya}, whereas Eq. (\ref{R_C/S}), which is valid for $m \lesssim m_{\rm Pl}$, represents the generalised Compton radius.

Though tentative, an identification of the form (\ref{alt_ident}) in the super-Planck mass regime would provide a concrete realisation of the BHUP correspondence, but not one based on modified de Broglie relations applied to fixed-background states \cite{Lake:2015pma,Lake:2016did,Lake:2016enn,Lake:2018hyv}, or on the inclusion of gravitational torsion \cite{Singh:2017wrb,Singh:2017ipg,Khanapurkar:2018tdd}, as in previous approaches. 

Equation (\ref{R_C/S-2}) is also consistent with gedanken experiment arguments previously presented in the literature. 
These suggest that there exist two irremoveable sources of error contributing to the position uncertainty of a black hole, whose linear dimension is estimated by observing its emitted Hawking radiation \cite{Maggiore:1993rv}.
The first, $\Delta x^{(1)} \simeq 2GM/c^2$, is simply the initial Schwarzschild radius, which corresponds to the position uncertainty of the emitted particle. 
The second is the change in the Schwarzschild radius due to the emission, $\Delta x^{(2)} \simeq 2G\Delta M/c^2$, where $\Delta M$ is the change in the black hole mass. 
Hence, setting $\Delta M \simeq m \simeq m_{\rm Pl}^2/M$, and assuming that the uncertainties add linearly, $\Delta X \simeq \Delta x^{(1)} + \Delta x^{(2)}$, we obtain an expression analogous to Eq. (\ref{R_C/S-2}). 
Following our previous convention, $\Delta X$, $\Delta x^{(1)}$ and $\Delta x^{(2)}$, discussed here, denote heuristic uncertainties, rather than well defined standard deviations. 

It is interesting to note, however, that an alternative line of reasoning allows us to derive a generalised position uncertainty for black holes which is analogous to Eq. (\ref{R_C/S-2}), but with the inequality in the opposite direction. 
In the classical-background theory, the hoop conjecture \cite{hoop} suggests the following criteria for the collapse of a self-gravitating quantum wave packet to form a black hole: 
\begin{eqnarray} \label{hoop_conj_fixed}
\Delta_{\psi}x^{\rm bh} \lesssim \frac{2G}{c^3}\Delta_{\psi}p^{\rm bh} \, . 
\end{eqnarray}
We may therefore conjecture that, in smeared-space, the equivalent condition is:
\begin{eqnarray} \label{hoop_conj_smeared}
\Delta_{\Psi}X^{\rm bh} = \sqrt{(\Delta_{\psi}x'^{\rm bh})^2 + 2l_{\rm Pl}^2}
\nonumber\\
\lesssim \frac{2G}{c^3}\Delta_{\psi}p'^{\rm bh} +  \frac{\hbar}{2\Delta_{\psi}p'^{\rm bh}} \, . 
\end{eqnarray}

A similar expression can be derived from gedanken experiment arguments analogous to those above by noting that, in fact, the first source of error in the position measurement of a black hole is given by $\Delta x^{(1)} \lesssim 2GM/c^2$. 
This follows from the fact that the black hole mass is localised within a radius not larger than its Schwarzschild radius, by the hoop conjecture.
(Operationally, we may say that the observed particles of Hawking radiation are emitted from within a linear region {\it not larger than} the Schwarzschild radius.)
Similarly, the second source of error is $\Delta x^{(2)} \simeq 2G\Delta M/c^2 \lesssim \hbar/(2Mc)$, where the final inequality follows from Eq. (\ref{alt_ident}).

If valid, Eq. (\ref{hoop_conj_smeared}) suggests a radically different form of generalised position uncertainty for black holes, vis-{\' a}-vis fundamental particles, as proposed in \cite{Lake:2015pma,Lake:2016did}. 
Namely, while the generalised Compton radius represents the minimum length-scale for the wave packet of a fundamental particle, beyond which pair-production occurs in place of further spatial localisation, the generalised event horizon represents the maximum length-scale for a quantum mechanical black hole, within which the wave function associated with its central mass is localised due to self-gravity.   

\subsection{Uncertainty relations in a finite universe} \label{FiniteU}

We shall now consider both lower and upper limits on the position and momentum uncertainties in the smeared-space formalism in more detail.  
These limits take into account the fact that, in every fixed-background geometry in the smeared-space superposition of geometries, the universe is of finite size in both position and momentum space.
These additional constraints lead to stricter bounds than previously discussed.

We first consider the case of canonical quantum theory in a finite-sized universe, where the maximum position uncertainty is given by the de Sitter length:
\begin{equation}
(\Delta_{\psi} x)_{\max} = l_{\mathrm{dS}} \, .
\end{equation}
Correspondingly, there exists a minimum momentum uncertainty that saturates the HUP,
\begin{equation}
(\Delta_{\psi} p)_{\min} = \frac{1}{2} m_{\mathrm{dS}} \, c \, , 
\end{equation}
which, as argued above Eq.~(\ref{sigma_g}), also sets the smearing scale for momentum space in our formalism: $\tilde{\sigma}_g := (\Delta_{\psi} p')_{\min}$. 
Here, the presence of a prime indicates a measurement in smeared-space, consistent with our previous notation.

Similarly, the boundary between the quantum (particle) and gravitational (black hole) regimes on the mass-radius diagram is given by the intersection of the Compton and Schwarzschild radii,
which implies the existence of a maximum mass for a fundamental particle, $m_{\max} = (1/\sqrt{2}) m_{\mathrm{Pl}}$ \cite{Carr:2014mya}.
This, in turn, gives rise to the minimum position uncertainty
\begin{equation}
(\Delta_{\psi} x)_{\min} = \frac{\hbar}{m_{\max} c} = \sqrt{2} l_{\mathrm{Pl}} \, .
\end{equation}
As also argued previously, this sets the smearing scale for position space: $\sigma_g := (\Delta_{\psi} x')_{\min}$.
By the HUP, the corresponding maximum momentum is
\begin{equation}
(\Delta_{\psi} p)_{\max} = \frac{1}{2\sqrt{2}} m_{\mathrm{Pl}} \, c \, . 
\end{equation}
Therefore, if the minimum position uncertainty were any smaller, the maximum energy density associated with the wave function of a quantum particle, $\mathcal{E}_{\psi} \simeq (\Delta_{\psi}p)_{\rm max}c/(\Delta_{\psi}x)_{\rm min}^3$, may exceed the Planck density, becoming large enough to induce collapse to a black hole.
Thus, the smeared-space position and momentum uncertainties are bounded, both from below and above, according to:
\begin{eqnarray} \label{EQ_XP_LIMITS}
2 l_{\mathrm{Pl}} \le & \Delta_{\Psi} X & \le \sqrt{l_{\mathrm{dS}}^2 + 2 l_{\mathrm{Pl}}^2} \, , 
\nonumber \\
\frac{1}{\sqrt{2}} m_{\mathrm{dS}} c \le & \Delta_{\Psi} P & \le \frac{1}{2} \sqrt{\frac{1}{2}m_{\mathrm{Pl}}^2 + m_{\mathrm{dS}}^2} c \, .
\end{eqnarray}

In the limit $l_{\mathrm{Pl}} \to 0$ ($m_{\mathrm{Pl}} \to \infty$) these bounds become
\begin{eqnarray}
0 \le & \Delta_{\psi} x & \le l_{\mathrm{dS}} \, , \nonumber \\
\frac{1}{\sqrt{2}} m_{\mathrm{dS}} c \le & \Delta_{\Psi} P & < \infty \, .
\end{eqnarray}
This corresponds to a scenario in which physical space remains classical, but in which there exists a finite maximum horizon distance, $r_{\rm H} = l_{\mathrm{dS}}$.
The existence of a finite horizon in physical space, in turn, implies the existence of a minimum possible momentum, $(\Delta_{\psi} p)_{\min} = \frac{1}{2} m_{\mathrm{dS}} c$, and, hence, of an innate (non-classical) smearing of momentum space. 
However, we may reverse this logic. 
Ergo, if there exists a finite minimum momentum, due to the innate smearing of momentum space, this gives rise to a minimum possible energy density, and, hence, to a maximum possible horizon in physical space \cite{Hobson:2006se,Spradlin:2001pw}.
Note that, as discussed below Eq. (\ref{beta-2}), the limit  $l_{\mathrm{Pl}} \to 0$ ($m_{\mathrm{Pl}} \to \infty$) is inconsistent in the smeared-space formalism. 
However, it is instructive to consider it as a hypothetical limit of the bounds (\ref{EQ_XP_LIMITS}), in order to develop our physical intuition.

Similarly, in the limit $l_{\mathrm{dS}} \to \infty$ ($m_{\mathrm{dS}} \to 0$), Eq. (\ref{EQ_XP_LIMITS}) yields
\begin{eqnarray}
2 l_{\mathrm{Pl}} \le & \Delta_{\Psi_g} X & < \infty \, , \nonumber \\
0 \le & \Delta_{\psi} p & \le \frac{1}{2\sqrt{2}} m_{\mathrm{Pl}} c \, .
\end{eqnarray}
This corresponds to scenario in which momentum space remains classical, but with a finite maximum horizon given by $\tilde{r}_{\rm H} = \frac{1}{2\sqrt{2}} m_{\mathrm{Pl}} c$, and position space is smeared. 
Again, this limit is inconsistent in the smeared-space formalism, but it is instructive to consider it, hypothetically, to aid our physical understanding.

The limits~(\ref{EQ_XP_LIMITS}) can now be understood intuitively. 
The lower bound on $\Delta_{\Psi}X$ arises from the Planck-scale smearing of spatial points, whereas the upper bound combines the limit due to a finite {\it classical} horizon with the Planck-scale smearing of the boundary points on the horizon itself.
Every point in the universe is Planck-scale smeared but the fluctuations in the interior region cancel out and only fluctuations of the boundary contribute to the upper limit on $\Delta_{\Psi} X$.
The existence of a finite classical horizon in position space can, in turn, be understood as a consequence of the minimum energy density implied by the innate smearing of points in momentum space.

Similarly, the lower bound on $\Delta_{\Psi}P$ arises from the innate de Sitter-scale smearing of momenta, whereas the upper bound combines this with the limit due to a finite classical horizon in momentum space.  
The latter can be understood as a consequence of the innate Planck-scale smearing of points in position space. 
This `momentum space horizon' marks the cut-off for the particle regime, beyond which the gravitational regime dominates.

The symmetry of the smeared-space model therefore implies {\it Planck-scale smearing of the de Sitter horizon} together with {\it de Sitter-scale smearing of the Planck point}, which marks the transition between the particle (quantum) and gravitational (classical) regimes \cite{Carr:2014mya}. 

Though it is beyond the scope of this paper to investigate these effects in detail, we note that our results suggest Planck-scale `smearing' of the gravitational singularity at the centre of a black hole, together with a concomitant smearing of the classical horizon. 
Thus, whatever their detailed implications, these findings are potentially relevant to models of black hole complementarity based on the holographic conjecture \cite{Bousso:2002ju}, to regular black hole models \cite{Hayward:2005gi}, and to the black hole information loss paradox \cite{Mathur:2009hf}.

We repeat that, within the formalism presented here, it is impossible to consider smearing either position or momentum space {\it alone}.
In this respect, the existence of a minimum vacuum energy, which may be identified with the existence of a cosmological constant term in the gravitational field equations, appears as an inevitable consequence of combining the quantum superposition principle with the existence of matter as the source of space-time curvature (gravity), implied by the principles of general relativity. 

Finally, we note that maximum and minimum bounds on position and momentum, together with generalised uncertainty relations of the form (\ref{GUP-final}) and (\ref{EUP-final}), can be implemented, naturally, by embedding our theory in the non-relativistic limit of the de Sitter geometry. 
We recall that canonical QM `lives' in flat Euclidean space, which is the non-relativistic limit of flat Minkowski space-time. 
The symmetry group of Minkowski space is the Poincar{\' e} group, which is the direct sum of the Lorentz group and the Galilean shift-isometry group. 
The Lorentz group comprises both Lorentz boosts and spatial rotations -- operations that leave the coordinate origin unchanged -- while the translation generators shift the origin of the coordinate axes. 

The symmetry group of de Sitter space is the de Sitter group, which is the unique one-parameter deformation of the Poincar{\' e} group \cite{deSittergroup}. 
In this, the Lorentz subgroup is preserved, but the distinction between rotations and translations is removed. 
Since the de Sitter space-time is closed, with constant positive curvature,
it may be foliated into space-like hyper-surfaces that exhibit spherical geometry \cite{Spradlin:2001pw}
\footnote{In fact, due to its extreme (maximal) symmetry, there exists no preferred space-like slicing in de Sitter space, which may be foliated by open, closed, and flat space-like hyper-surfaces. In all three cases, these exists a maximum horizon, $l_{\rm dS}$. However, in the non-relativistic limit, the closed-surface slicing appears more natural, since the finite size of the space-like $3$-spheres naturally mimics the finite horizon distance of the relativistic solution.}. 
A `translation' at a given point ${\vec x}$ may therefore be viewed as a rotation about an axis, centred on some other point ${\vec x}{\, '} \neq {\vec x}$.

Hence, we may construct a mathematically well defined theory in which $(\Delta_{\psi} x)_{\max} = l_{\mathrm{dS}}$ by placing the wave function $\psi(\vec{x})$ on a Riemannian background, corresponding to the non-relativistic limit of de Sitter space. 
The corresponding symmetry group is given by the Wigner-In{\"o}n{\"u} contraction of the de Sitter group and is known as the Newton-Hooke group \cite{Newton-Hooke:book,Guo:2004yq}. 
It is analogous to the Galilean group of flat Euclidean space, since it preserves Galilean boost invariance. 
However, as in the closed-surface slices of relativistic de Sitter space, the distinction between translations and rotations is removed due to the spherical symmetry of the (spatial) background geometry.

Thus, by identifying the momentum operator of the fixed-background theory $\hat{\vec{p}}$ with the `translation' generator of the Newton-Hooke group, with deformation parameter $\Lambda = 1/l_{\rm dS}^2$, one should obtain the limits $(\Delta_{\psi} p)_{\min} = (1/2)m_{\mathrm{dS}}c$, $(\Delta_{\psi} x)_{\max} = l_{\mathrm{dS}}$, as required. 
Similarly, we may obtain $(\Delta_{\psi} x)_{\min} \simeq l_{\mathrm{Pl}}$, $(\Delta_{\psi} p)_{\max} \simeq m_{\mathrm{Pl}}c$ by identifying the position operator $\hat{\vec{x}}$ with the Newton-Hooke `translation' generator in momentum space, with deformation parameter $\sim 1/(m_{\rm Pl}^2c^2)$.
In other words, we may implement both maximum and minimum position and momentum uncertainties by imposing (non-relativistic) de Sitter geometry on both the position and momentum space sub-manifolds of the fixed-background phase space. 

A recent study of the Heisenberg uncertainty principle, for wave functions defined on Riemannian 3-manifolds of constant curvature in canonical QM, supports the conclusions reached above.  
In \cite{Schurmann:2018yuz}, it was shown that the minimum momentum uncertainty is related to the curvature of physical space, $K_{x}$, via $(\Delta_{\psi}p)_{\rm min} = \hbar\sqrt{K_{x}}$. 
Clearly, in our case, $K_{x} = 1/l_{\rm dS}^2$.  
Similar considerations fix $(\Delta_{\psi}x)_{\rm min} = \hbar\sqrt{K_{p}}$, where $K_{p}$ is the curvature of momentum space. 
In our model, $K_{p} \simeq 1/(m_{\rm Pl}^2c^2)$.

\section{Conclusions} \label{CONCLUSIONS}

\subsection{Summary} \label{SUMMARY}

We introduced a quantum formalism capable of describing superpositions of classical geometries.
In this formalism, a point $\vec{x}$ in a classical background geometry is associated with a ket $\ket{\vec{x}}$ in a Hilbert space, and can make a coherent transition to any other point $\vec{x}{\, '}$. 
The transition is characterised by a quantum probability amplitude, $g(\vec{x}{\, '}-\vec{x})$. 
We argued that, in $d$ spatial dimensions, the most straightforward way of incorporating this possibility is via a quantum state with an additional $d$ degrees of freedom.

In the position space representation, the new degrees of freedom are represented by the vector $\vec{x}{\, '}$ and the function $g(\vec{x}{\, '}-\vec{x})$ is referred to as the `smearing function' for real space. 
In the momentum space representation, the new degrees of freedom are represented by the vector $\vec{p}{\, '}$ and the momentum space smearing function, $\tilde{g}_{\beta}(\vec{p}{\, '}-\vec{p})$, represents the quantum probability amplitude for the transition $\vec{p} \mapsto \vec{p}{\, '}$.
We naturally identified the width of the position space smearing function, in each orthogonal linear direction $x^{i}$, with the $D$-dimensional Planck length, where $D = d + 1$ is the number of space-time dimensions, i.e., $\sigma_g^{i} \simeq l_{\rm Pl}$. 
In addition, we argued that scale of momentum space smearing should be set by the de Sitter scale, giving $\tilde{\sigma}_{gi} \simeq \hbar/l_{\rm dS}$, where $l_{\rm dS} = \sqrt{d/\Lambda_{D}}$ is the de Sitter radius. 
In $(3+1)$-dimensional space-time (our observable universe), $l_{\rm dS} = \sqrt{3/\Lambda}$, where $\Lambda \simeq 10^{-56} \, {\rm cm}^{-2}$ is the observed value of cosmological constant, and $l_{\rm dS} \simeq 10^{28}$ cm is comparable to the present day horizon radius. 
However, although we defined the new formalism in an arbitrary number of spatial dimensions, $d$, we made no attempt to `smear' $D$-dimensional space-time, and restricted our attention to the non-relativistic regime.

Thus, an interesting aspect of the smeared-space formalism is that consistency {\it requires} the existence of a nonzero dark energy density, i.e., $\rho_{\Lambda} = \Lambda c^2/(8\pi G)$ ($\Lambda >0$) for $D=3+1$. 
This follows automatically from the fact that the smearing of position space, which introduces a minimum length $\sim l_{\rm Pl}$, implies a concomitant smearing of momentum space, thus introducing a minimum momentum $\sim l_{\rm dS}$.  
From these considerations, it follows that $g(\vec{x}{\, '}-\vec{x})$ and $\tilde{g}_{\beta}(\vec{p}{\, '}-\vec{p})$ are related by the Fourier transforms, performed at the scale $\beta := (2/d)\sigma_g^{i}\tilde{\sigma}_{gi}$, rather than $\hbar$, which sets the transformation scale between $\psi(\vec{x})$ and $\tilde{\psi}(\vec{p})$ in canonical QM. 
For $d=3$, $\beta \simeq \hbar \sqrt{\rho_{\Lambda}/\rho_{\rm Pl}} \simeq \hbar \times \mathcal{O}(10^{-61})$.

This, in turn, implies a modification of the canonical de Broglie relation between the wave vector and momentum, such that the physically observable momentum ($\vec{p}{\, '}$) is given by $\vec{p}{\, '} = \hbar \vec{k} + \beta(\vec{k}'-\vec{k})$. 
Here, $\vec{k}$ denotes the usual de Broglie wave vector, which is associated with the momentum eigenstate of a quantum particle on a classical background space. 
Heuristically, the term $\beta (\vec{k}'-\vec{k})$ can be understood as the possible `kick', given to a point on the plane-wave, due to the transition $\vec{x} \mapsto \vec{x}{\, '}$ in the smeared geometry. 

These considerations imply a kind of `wave-point' duality, analogous to wave-particle duality in canonical QM, which gives rise to an uncertainty relation for spatial `points', $\Delta_{g} x'^{i} \, \Delta_{g} p'_{j} \geq (\beta/2) \, \delta^{i}{}_{j}$ (\ref{EQ_BETA_SCALE}). 
The inequality is saturated when the smearing functions are chosen to be Gaussians, which justifies our definition of the transformation scale: $\beta := (2/d)\sigma_g^{i}\tilde{\sigma}_{gi}$. 
Nonetheless, the relation holds for arbitrary $g$, yielding $\Delta_{g} x'^{i} \, \Delta_{g} p'_{j} > (\beta/2) \, \delta^{i}{}_{j}$ for non-Gaussian smearing.

Though a detailed investigation of the wave-point uncertainty relation lies beyond the scope of the present work, we note that, na{\" i}vely, it implies that the momentum of a graviton (assuming such a particle can be consistently defined in a quantum gravitational framework) should be given by $p' \sim \beta/\lambda'$, where $\lambda'$ is its observable wavelength, rather than $p \sim h/\lambda$, as commonly assumed.  
In other words, our formalism suggests that the quantum mechanics of space is characterised by a radically different scale ($\beta$) than that which characterises the quantum mechanics of matter ($\hbar$). 
Furthermore, if the gravitational coupling is renormalised at higher energies, i.e., if $G$ is a running coupling like the couplings of the standard model fields, then the quantisation scale $\beta$ is not fixed, but is also energy-dependent. 

The modified de Broglie relation and choice of smearing functions uniquely determine the smeared-space formalism.
Within this formalism, it was shown how to calculate the statistics of generalised position and momentum measurements, with emphasis on the scenario in which $g(\vec{x}{\, '}-\vec{x})$ and $\tilde{g}_{\beta}(\vec{p}{\, '}-\vec{p})$ are Gaussian functions,
centred at $\vec{x}{\, '} = \vec{x}$ and $\vec{p}{\, '} = \vec{p}$, respectively. 
The resulting generalised observables naturally incorporate fundamental limits on the precision of position and momentum measurements, given, respectively, by $\sigma_g^{i}$ and $\tilde{\sigma}_{gi}$. 
For particles of quantum matter in the smeared-space background, this implies that both the GUP and EUP hold, independently, in practically all situations of physical interest.   
They may also be combined to give a unified uncertainty relation that is symmetric in position and momentum, Eq. (\ref{X_g_P_X_g_uncertainty-2}).
This is the first key result of this paper.

Next, we considered the position and momentum space representations in smeared-space wave mechanics, and derived a generalised Schr{\" o}dinger equation for particles propagating in the smeared geometry. 
The associated Heisenberg equation was also obtained. 
As in canonical QM, the time-derivative of a general observable $\hat{O}(t)$ was found to be proportional to the commutator $[\hat{H},\hat{O}]$, suggesting a canonical-type quantisation scheme for the smeared-space model: 
$\left\{O_1,O_2\right\} \mapsto {\rm const.} \times i \, [\hat{O}_1,\hat{O}_2]$. 
The modified position-momentum commutator in the smeared-space theory was calculated as $[\hat{X}^{i},\hat{P}_{j}] = i(\hbar + \beta) \delta^{i}{}_{j}\ {\bf\hat{\mathbb{I}}}$ (\ref{[X,P]}), suggesting ${\rm const.} = \hbar + \beta$.
Equation (\ref{[X,P]}) is the second key result of this paper. 

We recall that generalised uncertainty relations derived from modified commutation relations, in which the position-momentum commutator is no longer proportional to the identity matrix (see Eq. (\ref{modified_commutator-1})), are believed to imply violation of the equivalence principle \cite{Tawfik:2015rva,Tawfik:2014zca}. 
This is true regardless of whether the modification arises from the modified symmetry group of physical space or from a modification of the canonical de Broglie relation, such that $\vec{p}(\vec{k})$ is a nonlinear function of $\vec{k}$. 
This remains a fundamental objection to their acceptance. 

In the first scenario, one is faced with an additional problem in the classical limit of the theory. 
Namely, implementing a canonical quantisation scheme and requiring the correspondence principle \cite{Rae} to hold implies an equivalent modification of the canonical Poisson brackets. 
This implies violation of Galilean invariance, even for macroscopic (classical) systems, and, hence, violation of Poincar{\' e} invariance in the relativistic limit. 
To date, no definitive observational evidence for the breaking of Poincar{\' e} invariance (including shift-invariance) has been obtained, though bounds on the symmetry-breaking parameters have been determined from a variety of experiments \cite{Tawfik:2015rva,Tawfik:2014zca,Hossenfelder:2012jw}. 
In the second scenario, one also encounters theoretical problems related to the nonlinearity of $p(k)$ in the relativistic regime. 
(Here, $p$ and $k$ denote the relativistic $4$-momentum and its corresponding wave number, respectively, though we neglect space-time indices for the sake of notational convenience.)
We now discuss the most serious of these problems, which concerns the construction of multi-particle states, noting that, in the smeared-space model, it does not occur.

When $p(k)$ is nonlinear, it is unclear whether one should require the physical momentum $p$, or wave number $k$ (also known as the pseudo-momentum), to transform under the Poincar{\'e} group. 
Choosing wave number as the Lorentz-invariant quantity, the Lorentz transformations become nonlinear functions of $k$. 
The transformation of the sum $k_1+k_2$ is then no longer equal to the sum of the transformations of $k_1$ and $k_2$, individually. 
Conversely, choosing $p$ as the Lorentz-invariant variable (which is physically more reasonable), the opposite is true, i.e., the transformation of the sum $p_1+p_2$ is no longer equal to the sum of the individual transformations of $p_1$ and $p_2$. 
Each case requires the definition of new nonlinear addition law, either for pseudo-momenta, or for physical momenta, respectively.  

In the latter, the new sum rule for physical momenta is frame-independent, by construction, but a new problem is created:  
if the nonlinear composition function has a maximum at the Planck momentum, then the sum of momenta will never exceed this maximum value. 
Although the Planck momentum is large, for fundamental particles with $m \ll m_{\rm Pl}$, it is small for macro-objects with masses $M \gg m_{\rm Pl}$, which may easily exceed it at non-relativistic velocities.  
(We recall that $m_{\rm Pl} \simeq 10^{-5}$ g.)
The problem of reproducing a sensible multi-particle limit when one chooses the physical momentum to transform under modified Lorentz transformations is thus known as the `soccer ball problem' \cite{Hossenfelder:2012jw}. 

In summary, since the introduction of nonlinear de Broglie relations in non-relativistic quantum theory necessarily implies nonlinear relations in the relativistic limit, Lorentz violation is unavoidable. 
Though this can be overcome, theoretically, by introducing an appropriate nonlinear addition law, this leads to new problems, and it is not clear whether sensible multi-particle
limits of such theories exist \cite{Hossenfelder:2012jw}. 
Therefore, it is not clear whether generalisations of canonical QM based on nonlinear {\it non-relativistic} de Broglie relations, $\vec{p}(\vec{k})$, admit sensible multi-particle limits either. 
Nonetheless, these form the basis of all current implementations of the GUP \cite{Tawfik:2015rva,Tawfik:2014zca}.   
However, in the smeared-space formalism, multi-particle states that are invariant under the usual Galilean symmetries of non-relativistic systems can be easily defined, as in Sec.~\ref{MULTI}. 
Hence, the `soccer ball problem' does not arise.

This is related to another, somewhat subtle result of the smeared-space model. 
We note that in the smeared-space formalism Poincar{\'e} symmetry is neither broken, as it is in loop quantum gravity (LQG) \cite{Ashtekar:2012np} and non-commutative geometry (NCG) \cite{Connes}, nor unbroken, as it is in string theory \cite{Kiritsis:2007zza}, but `smeared'. 
Specifically, the non-relativistic limit of Poincar{\'e} invariance (i.e., Galilean invariance) is preserved in the extended phase space, including the additional degrees of freedom labeled by $\vec{x}{\, '}$ or $\vec{p}{\, '}$. 
This corresponds to the smeared-state picture. 
By contrast, in the smeared-operator picture, operators act on the fixed-background state $\ket{\psi}$ and superpositions of isometries, weighted by the functions $g(\vec{x}{\, '}-\vec{x})$ or $\tilde{g}_{\beta}(\vec{p}{\, '}-\vec{p})$ depending on which representation we choose, are encoded in the operators themselves. 

Thus, in the smeared-space model, the usual Galilean isometries and the standard non-relativistic de Broglie relation $\vec{p} = \hbar \vec{k}$ hold in each classical background in the smeared superposition of geometries. 
Hence, the equivalence principle also holds in each individual geometry. 
In this way, our approach overcomes a serious theoretical objection to the implementation of generalised uncertainty relations based on modified commutators. 

Finally, the implications of the smeared-space model for cosmology and black hole physics were also considered. 
The optimum values of the position and momentum uncertainties, which minimise the lower bound on the right-hand side of the unified uncertainty relation (\ref{X_g_P_X_g_uncertainty-2}) were determined, yielding $(\Delta_{\psi}x')_{\rm opt} \simeq l_{\Lambda}$, $(\Delta_{\psi}p')_{\rm opt} \simeq m_{\Lambda}c$, where $l_{\Lambda} \simeq \sqrt{l_{\rm Pl}l_{\rm dS}} \simeq 0.1$ mm and $m_{\Lambda} \simeq \sqrt{m_{\rm Pl}m_{\rm dS}} \simeq 10^{-3} \, {\rm eV/c^2}$ ($m_{\rm dS} = \hbar/l_{\rm dS}$) in $(3+1)$ dimensions. 
These values are of particular relevance to cosmology, since $\rho_{\Lambda} \simeq (3/4\pi)m_{\Lambda}/l_{\Lambda}^3$. 
In other words, when the wave function of a quantum particle saturates the inequality in the smeared-space uncertainty relations, (\ref{GUR_X})-(\ref{GUR_P}), its associated energy density is of the order of the dark energy density. 
This suggests a granular model of dark energy, in which the dark energy density is approximately constant on large scales, but exhibits spatial variations on a length-scale of order $\sim 0.1$ mm. 
Tentative observational evidence for periodic variation of the gravitational field strength on this scale \cite{Perivolaropoulos:2016ucs,Antoniou:2017mhs} was also discussed. 

Though, strictly, the smeared-space formalism is valid only for $m \lesssim m_{\rm Pl}$, a physical argument was constructed that allowed us to tentatively extend the GUP into the regime $m \gtrsim m_{\rm Pl}$.
By considering the observation of emitted particles of Hawking radiation, the smeared-space GUP was used to obtain bounds on the localisation of both fundamental particles and black holes, yielding expressions for the generalised Compton wavelength and generalised event horizon, respectively. 
This yields a concrete implementation of the black hole uncertainty principle (BHUP) correspondence \cite{Carr:2014mya}, since the two expressions form a unified curve on the mass-radius diagram which is valid for all mass- and length-scales. 

\subsection{Future work} \label{FUTURE}

We conclude with a few comments on the extension of the present approach:
\begin{itemize}

\item 
Since the `smearing' probability amplitude $g(\vec{x}{\, '}-\vec{x})$ is only a function of the distance between the points that make the transition, it is assumed that all classical points are smeared in the same way.
While this is realistic in spaces that are approximately flat, i.e., on which non-relativistic particles with masses $m \ll m_{\rm Pl}$ propagate, a more complete treatment should include the gravitational field generated by quantum particles themselves. 

In other words, the present approach neglects the back-reaction of the energy density, associated with the wave function $\psi(\vec{x})$, on the background geometry.  
It should therefore be extended to include the effects of the Newtonian gravitational potential. 
This, in turn, is equivalent to extending the smearing procedure from flat Euclidean space to general Riemannian spaces, since Newtonian gravity may be reinterpreted as theory of spatial curvature with an absolute time parameter, see \cite{Hansen:2018ofj,Banerjee:2018gqz}.  

\item 
Similarly, we may extend the smearing procedure to flat Minkowski space. 
In this scenario, the self-gravity of quantum particles would still be neglected, but special-relativistic effects may be taken into account. 
This corresponds to the development of quantum field theory on a smeared space-time background and, ultimately, should include a `smeared' version of the standard model of particle physics. 

An important open question for such a model is whether space-time symmetries can be smeared without a concomitant smearing of gauge invariance. 
Yet another is the generalisation of the path-integral procedure to include an infinite sum over background geometries. 
In such an extension we expect there to be an infinite number of paths corresponding to each Feynman path in the fixed Minkowski space-time, i.e., a path-integral within a path-integral, which sums over the additional geometric degrees of freedom. 

\item
Were the preceding two projects to be successfully completed, we would be provided with two possible routes by which to attack the fundamental problem of quantum gravity, namely, the description of quantum superpositions of pseudo-Riemannian geometries (i.e., curved space-times). 
This may be regarded as the ultimate goal of our research.  

\item 
We note that our approach is non-perturbative, in the sense that we do {\it not} quantise perturbations around a fixed classical background geometry.  
Instead, the Euclidean background arises as the most probable state in the quantum superposition of geometries, and all possibilities are included in the sum over amplitudes $g(\vec{x}{\, '}-\vec{x})$. 
The possible connections between the smeared-space theory and non-perturbative results in quantum field theory, high-energy physics, and the existing quantum gravity literature should therefore be explored. 

\item 
Since the smearing function $g(\vec{x}{\, '}-\vec{x})$ is interpreted as a property of quantum mechanical space, as opposed to $\psi(\vec{x})$, which describes matter on a classical background, the smeared-space wave function $\Psi(\vec{x},\vec{x}{\, '}) := g(\vec{x}{\, '}-\vec{x})\psi(\vec{x})$ entangles matter and geometry.
It may be hoped, therefore, that the model gives rise to a concrete realisation of the gravity-matter entanglement hypothesis put forward in \cite{Kay:2018mxr,Kay:2007rx}, though further work is needed to determine how this relates to the Newtonian gravitational potential.

\item
In more than one spatial dimension, the smeared-space formalism also gives rise to uncertainty relations of the form $\Delta_{\Psi}X^{i}\Delta_{\Psi}X^{j} \gtrsim l_{\rm Pl}^2$ and $\Delta_{\Psi}P_{i}\Delta_{\Psi}P_{j} \gtrsim m_{\rm dS}^2c^2$, for all $i, \, j$ corresponding to linear coordinate directions. 
Such uncertainty relations arise naturally in NCG \cite{Connes}. 
However, here, we obtain them in the presence of {\it commuting} coordinates: $[\hat{X}^{i},\hat{X}^{j}] = 0$, $[\hat{P}_{i},\hat{P}_{j}] = 0$. 
This is related to the subtle point, mentioned in Sec. \ref{SUMMARY}, that in the smeared-space formalism space is not discretised, as it is in LQG \cite{Ashtekar:2012np}, and in some models of NCG \cite{Kauffman}. 
Instead, the smearing procedure represents a quantisation of the continuum, i.e., homogeneous real space $\mathbb{R}^d$. 
That said, it is clear that the commutative nature of the generalised position and momentum operators stems directly from the assumption that position and momentum space are smeared {\it independently}. 
Relaxing this assumption, and allowing transitions between points in the position and momentum subspaces of the classical phase space, e.g., such that $x^{i} \rightarrow x'^{i} + \theta^{ik}p'_{k}$, $p_{j} \rightarrow p'_{j} + \tau_{jk}x'^{k}$, it is not difficult to see how a generalised model could give rise to commutators of the form $[\hat{X}^{i},\hat{X}^{j}] \sim \theta^{ij} \, {\bf\hat{\mathbb{I}}}$ and $[\hat{P}_{i},\hat{P}_{j}] \sim \tau_{ij} \, {\bf\hat{\mathbb{I}}}$, as well as $[\hat{X}^{i},\hat{P}_{j}] \sim \delta^{i}{}_{j} \, {\bf\hat{\mathbb{I}}}$. 
In particular, we may expect to realise the Seiberg-Witten map \cite{Bertolami:2015yga} for an appropriate choice of model parameters.

\item 
Finally, we note that the generalisation of the Feynman path-integral corresponds to assigning a weight to each path $\vec{x}{\, '}(\vec{x})$ in the extended phase space, illustrated heuristically in Fig.~\ref{FIG_PHASESPACE}. 
In Sec.~\ref{SEC_FORMALISM}, we outlined how the induced metric on $\vec{x}{\, '}(\vec{x})$ could be obtained as push-forward from the metric on $(\vec{x},\vec{x}{\, '})$-plane.   
We now outline physical reasons for believing that the bulk-space metric is the $(d+d)$-dimensional Minkowski metric, as previously claimed. 

Take, for example, the line $(\vec{x},\vec{0})$, which corresponds to the `original' un-smeared background geometry. 
The point $(\vec{a},\vec{a}{\, '})$ in the $(\vec{x},\vec{x}{\, '})$-plane corresponds to the non-local influence of the point $\vec{a}{\, '}$ on $\vec{a}$. 
By symmetry, it also refers to the non-local influence of $\vec{a}$ on $\vec{a}{\, '}$. 
Hence, both the point $(\vec{a},\vec{a}{\, '})$ and the point $(\vec{a}-\vec{a}{\, '},\vec{0})$ refer to the same physical process, i.e., the transition $\vec{a} \mapsto \vec{a}{\, '}$. 
(Each is associated with the same quantum probability amplitude, $g(\vec{a}{\, '}-\vec{a})$, assuming that $g(\vec{x})$ is symmetric about $\vec{x}=\vec{0}$.)
Thus, the distance between the points $(\vec{a},\vec{a}{\, '})$ and $(\vec{a}-\vec{a}{\, '},\vec{0})$ on the $(\vec{x},\vec{x}{\, '})$-plane should be zero, for any $\vec{a}$, $\vec{a}{\, '}$, which is most easily achieved by imposing the metric ${\rm d}s^2 = -{\rm d}\vec{x}{\, '}^2 + {\rm d}\vec{x}^2$.

However, in order to calculate the probability associated with a given geometry, we must sum over points on the curve $\vec{x}{\, '}(\vec{x})$, taking into account their weighted amplitudes $g(\vec{x}{\, '}-\vec{x})$. 
In so doing, we must generalise the definition of a classical metric to include such weights. 
To this end, we note that weighted `metrics' have been studied in the mathematical literature on probability theory.    
Though not strictly a metric in the usual sense, a probability-weighted distance measure has already been defined, and is known as the {\L}ukaszyk-Karmowski metric \cite{LK_metric}.  
Our task, therefore, is to generalise this probability-weighted measure to include amplitude-weighted sums, {\`a} la canonical path-integral techniques. 
However, here, the associated amplitudes are not interpreted as the wave functions of quantum particles on classical backgrounds. 
Instead, they represent the weights associated with spatial points in an entangled superposition of geometries, represented by a higher-dimensional phase space. 

We emphasise that, even if the metric on the $2d$-dimensional hyper-plane $(\vec{x},\vec{x}{\, '})$ contains time-like directions, observable $d$-dimensional sub-manifolds represent {\it spatial} geometries. 
Strictly, if this is indeed the case, we must introduce an additional factor of $(-1)^d$ into the volume elements ${\rm d}^d\vec{x}{\rm d}^d\vec{x}{\, '}$ and ${\rm d}^d\vec{p}{\rm d}^d\vec{p}{\, '}$, used in the position and momentum space representations of the smeared-state $\ket{\Psi}$, as well as the associated definitions of $\hat{X}^i$ and $\hat{P}_j$. 
This is because the $(d+d)$-dimensional Minkowski metric is the direct sum of two $d$-dimensional Euclidean metrics with opposite signatures, i.e., $\eta_{IJ}(x,x') = (-E_{ij}(x')) \oplus E_{ij}(x)$, where $I,J \in \left\{1,2, \dots 2d \right\}$. 
The determinant then factorises such that $\det \eta_{IJ}(x,x') = (-1)^d \det E_{ij}(x) \det E_{ij}(x')$. 
Hence, Eqs. (\ref{EQ_PSIG}) and (\ref{EQ_PSIG*}), and all  associated definitions of observables in the smeared-space formalism, may be corrected simply by including an appropriate sign factor in the Jacobian.

\end{itemize}

\acknowledgments

This work is supported by Singapore Ministry of Education Academic Research Fund Tier 1 Project No. RG106/17. 
SDL was supported by the Natural Science Foundation of Guangdong Province, grant no. 2016A030313313.

\renewcommand{\theequation}{A-\arabic{equation}}
\setcounter{equation}{0}  
\section*{Appendix: An effective model in the classical background} 

In this Appendix, we show that the modified uncertainty relations derived from the smeared-space formalism can also be obtained in an effective model, where position and momentum measurements in {\it canonical quantum theory} are imprecise and described by POVMs, rather than perfect projective measurements.
We note, however, that there exists a fundamental physical difference between the smeared-space formalism and the POVM approach, namely that, in the latter, measurements are performed on the canonical quantum state, $\psi(\vec{x})$ or $\tilde{\psi}_{\hbar}(\vec{p})$, which is defined on a fixed classical background geometry.
As such, there is nothing {\it fundamental} about the finite-precision limits, which we may again label $\sigma_g^{i}$ and $\tilde{\sigma}_{gi}$, that arise in this model. 

In particular, we may imagine preparing an ensemble of states, using a finite-precision measuring device corresponding to our POVM, each with position uncertainty $\Delta_\psi x^{i} \simeq \sigma_g^{i}$.
We are then free to superpose these POVM-prepared states in such a way as to create a state with width $\Delta_\psi x^{i} \ll \sigma_g^{i}$.
Similar considerations apply to states prepared with momentum uncertainty $\Delta_\psi p_{i} \simeq \tilde{\sigma}_{gi}$.
In short, since the states $\psi(\vec{x})$ and $\tilde{\psi}_{\hbar}(\vec{p})$ are canonical quantum states, defined on a fixed classical background, there are no fundamental limitations to their position and momentum spreads.
Instead, the uncertainty relations we now derive hold if space-time is perfectly sharp but the measuring devices we use nonetheless have finite precision.

Let us begin by replacing the usual position-measurement operator, $\hat{x}^{i}$, with measurement operators defined as
\begin{equation} \label{E_x}
\hat{E}_{\vec{x}} := \int g(\vec{x}{\, '} - \vec{x}) \proj{\vec{x}{\, '}} {\rm d}^d\vec{x}{\, '} \, .
\end{equation}
These give rise to POVM elements, $\hat{E}_{\vec{x}}^\dagger \hat{E}_{\vec{x}} \ge 0$, satisfying $\int \hat{E}_{\vec{x}}^\dagger \hat{E}_{\vec{x}} {\rm d}^d\vec{x} = \hat{\openone}$, as required.
We emphasise that, in this scenario, there is no extra degree of freedom, since $\vec{x}{\, '}$ is simply a dummy variable in the integrand.
Thus, Eq. (\ref{E_x}) defines a standard POVM in canonical $d$-dimensional quantum mechanics.

Finite-precision measurements, conducted on an arbitrary state $\ket{\psi}$, then give rise to the following moments: 
\begin{eqnarray}
\langle E_{\vec{x}} \rangle_\psi & = & \langle \vec{x} \rangle_{g} + \langle \vec{x} \rangle_\psi \, , 
\nonumber\\
\langle E_{\vec{x}}^2 \rangle_\psi  & = & \langle \vec{x}^2 \rangle_{g} + \langle \vec{x}^2 \rangle_\psi \, , 
\end{eqnarray}
where $\langle \vec{x}^n \rangle_{g} := \int \vec{x}^n \, |g(\vec{x})|^2 \, {\rm d}^d\vec{x}$. 
Assuming that $|g(\vec{x}{\, '}-\vec{x})|^2$ is a normalised function with width $\sigma_g^{i}$, in each coordinate direction $x^{i}$,  and is centred at $\vec{x}{\, '}=\vec{x}$ (i.e., so that $\langle \vec{x} \rangle_{g} = 0$), 
the corresponding variance is given by:
\begin{eqnarray}
(\Delta_\psi E_{\vec{x}})^2 = (\Delta_\psi \vec{x})^2 + \vec{\sigma}_g^2 \, ,
\end{eqnarray}
where $\vec{\sigma}_g := \sigma_g^{i}{\bf e}_{i}$.

In the same manner, one introduces imperfect momentum measurement via the operators
\begin{equation} \label{E_p}
\hat{E}_{\vec{p}} := \int \tilde{g}(\vec{p}{\, '} - \vec{p}) \proj{\vec{p}{\, '}} {\rm d}^d\vec{p}{\, '} \, .
\end{equation}
Assuming that the function $|\tilde{g}(\vec{p}{\, '}-\vec{p})|^2$ has width $\tilde{\sigma}_{gi}$, in each coordinate direction, and is centred at $\vec{p}{\, '} = \vec{p}$ (i.e., so that $\langle \vec{p} \rangle_{g} = 0$), we then have:
\begin{eqnarray}
(\Delta_\psi E_{\vec{p}})^2 = (\Delta_\psi \vec{p})^2 + \vec{\tilde{\sigma}}_g^2 \, ,
\end{eqnarray}
where $\vec{\tilde{\sigma}}_g := \tilde{\sigma}_{gi}{\bf e}^{i}$.

Clearly, these two variances are of the same general form as the ones derived in the smeared-space formalism, and hence give rise to uncertainty relations of the same form as the GUP and EUP, respectively. 
However, we note that, in this case, the finite-precision uncertainties $\sigma_g^{i}$ and $\tilde{\sigma}_{gi}$ are not intrinsically related.
Although, mathematically, $|\tilde{g}(\vec{p}{\, '}-\vec{p})|^2$ may be written as the Fourier transform of $|g(\vec{x}{\, '}-\vec{x})|^2$ (performed at some arbitrary scale) this does not imply a concomitant modification of the canonical de Broglie relations, or of the ideal projective measurement operators, $\hat{\vec{x}}$, $\hat{\vec{p}}$ and $\hat{H} = \hat{p}^2/(2m)$, etc.
Hence, it does not imply a modification of the standard Schr{\"o}dinger equation, or of the basic conceptual framework of canonical QM.


\end{document}